\documentclass[aps,pra,preprint,groupedaddress]{revtex4}
\usepackage{epsfig,amsmath,graphicx,amssymb}
\usepackage{graphicx}
\usepackage{dcolumn}
\usepackage{bm}
\usepackage{CJK}

\def\be{\beta}

\def\be{\begin{equation}}
\def\ee{\end{equation}}
\def\bea{\begin{eqnarray}}
\def\eea{\end{eqnarray}}
\def\bes{\begin{subequations}}
\def\ees{\end{subequations}}

\begin{document}

\title{Collective oscillation modes of a superfluid Bose-Fermi mixture}

\author{Wen Wen$^{1,2}$}
\email[]{wenwen0emma@163.com}
\author{Ying Wang$^{3}$}
\author{Jianyong Wang$^{1}$}
\affiliation{$^{1}$ Department of Mathematics and Physics, Hohai University, Changzhou 213022, China\\
              $^{2}$ College of Science, Hohai University, Nanjing 210098, China\\
              $^{3}$ School of Science, Jiangsu University of Science and Technology, Zhengjiang 212003, China}

\date{\today}

\begin{abstract}
In this work, we present a theoretical study for the collective oscillation modes, i.e. quadrupole, radial and axial mode, of a mixture of
Bose and Fermi superfluids in the crossover from a Bardeen-Cooper-Schrieffer (BCS) superfluid to a molecular Bose-Einstein condensate (BEC)
in harmonic trapping potentials with cylindrical symmetry of experimental interest. To this end, we start from
the coupled superfluid hydrodynamic equations for the dynamics of Bose-Fermi superfluid mixtures and use the scaling theory that
has been developed for a coupled system. The collective oscillation modes of Bose-Fermi superfluid mixtures
are found to crucially depend on the overlap integrals of the spatial derivations of density profiles of the Bose and Fermi superfluids
at equilibrium. We not only present the explicit expressions for
the overlap density integrals, as well as the frequencies of the collective modes provided that the effective Bose-Fermi coupling is weak,
but also test the valid regimes of the analytical approximations by numerical calculations in realistic experimental conditions. In the presence of a repulsive Bose-Fermi interaction, we find that the frequencies
of the three collective modes of the Bose and Fermi superfluids are all upshifted, and the change speeds of the frequency shifts in the BCS-BEC crossover can characterize the different groundstate phases of
the Bose-Fermi superfluid mixtures for different trap geometries.
\end{abstract}

\maketitle
\section{introduction}
The recent experimental realization of a mixture of Bose and Fermi superfluids in ultracold
atoms has generated much interest in the properties of this new state of matter
\cite{fer2014,del2015,yao2016,takuya2017,roy2017,yuping2018},
and led to a surge of theoretical activity \cite{oza2014,ren2014,cui2014,zheng2014,kin2015,zi2018,chevy2015,nes2017,
tyl2016,jiang2017,pan2017,ling2014}. Collective excitations that characterise a system's response
to small perturbations constitute one of the main sources of information for understanding
the physics of many-body systems \cite{dal1999,gio2008}. In the last two decades, collective modes have been extensively investigated
to understand the properties of atomic gases in various systems
\cite{edw1996,sstri1996,vic1996,hei2004,hui2004,gari2004,ast2005,bul2005,yin2006,adhk2010,pu1998,busch1997,rod2004,miy2000,liu2003,mar2005,mar2009,may2013},
including bosons \cite{bosec,che2002}, fermions \cite{kin2004,fermc},
and multi-components, such as spin-orbit-coupled bosons \cite{zhang2012} and fermions \cite{zhang2018},
Bose-Bose mixtures \cite{bbmc} and degenerate Bose-Fermi mixtures \cite{huang2018,fuk2009}.

The Bose-Fermi superfluid mixture \cite{fer2014,del2015,yao2016,takuya2017,roy2017,yuping2018},
which is different from other coupled systems obtained previously \cite{zhang2012,zhang2018,bbmc,huang2018,fuk2009},
can provide a unique setting for studying and understanding the properties of interacting quantum systems belonging to different
quantum statistics. The fermion-fermion interactions can be widely tuned with a magnetic field Feshbach resonance.
For the strongly-repulsive interaction, it is a mixture of two BEC, in which one made of atoms and the other of molecules.
For the weakly-attractive interaction it is a mixture of a Bose superfluid
with a BCS-type superfluid. Such strong interactions are difficult to generate with bosonic atoms.
However, three body losses and recombination processes are significantly lower with fermions due to the Pauli principle.
The lifetime of a mixture of Bose and Fermi superfluids has been demonstrated experimentally to be in the order of a few seconds \cite{fer2014}.
Such stability allows us to observe this mixture oscillating back and forth in harmonic traps over numerous periods
without visible damping, and study how the Bose-Fermi interaction affects the dipole modes of Bose
and Fermi superfluids. The long-lived center-of-mass oscillations have been realized in the Bose-Fermi superfluid mixtures of
{$^7$}Li-{$^6$}Li \cite{fer2014,del2015}, {$^{41}$}K-{$^6$}Li \cite{yuping2018}, and {$^{174}$}Yb-{$^6$}Li \cite{roy2017}, and the Bose-Fermi interaction gives
rise to a rich behavior. In the presence of a repulsive Bose-Fermi interaction, the frequencies of the dipole oscillations
of the Bose and Fermi superfluids are both downshifted in the weakly confined direction \cite{fer2014,roy2017,yuping2018}, and
the frequency shifts increase monotonically from the BCS side to the BEC side \cite{wen2017,wen2018}. In contrast,
the frequency in the tight confinement for the Bose superfluid is upshifted, whereas the frequency for the Fermi superfluid
is still downshifted \cite{yuping2018}. The frequency shifts show non-monotonic and resonantlike behaviors in both directions
around the BCS side \cite{yuping2018}, which may be originated from the effects of fermionic pairs breaking \cite{ren2018}.

It is naturally to ask how Bose-Fermi interaction affects the collective
modes of Bose and Fermi superfluids in the BCS-BEC crossover, which is of great interest recently.  However, a theoretical
study for the collective oscillation modes of Bose-Fermi superfluid mixtures in a realistic experimental situation
is highly nontrivial. In this work, to study the quadrupole, radial and axial modes in cylindrically symmetric traps,
we start from the coupled superfluid hydrodynamic equations describing the dynamics of
the Bose-Fermi superfluid mixtures. The scaling method for coupled systems is then applied, and the eigenvalue equations
for the coupled collective modes are obtained, which are crucially sensitive to
the overlap integrals of the spatial derivations of the Bose and Fermi densities at groundstate. To present the
explicit expressions for these integrals, we use a perturbative approximation for the coupled Bose and Fermi density profiles
and in the overlap region the Fermi density is replaced by the value in the trap center.
The analytical results for the frequencies of collective oscillation modes are calculated in realistic experimental parameters \cite{yuping2018},
and the valid regimes of the analytical approximations are confirmed by the numerical calculations.
In the presence of a repulsive Bose-Fermi interaction, we find that the frequencies for the collective modes
of the Bose and Fermi superfluids are all upshifted, and the frequency shifts for the Fermi superfluid are smaller than the
bosons, due to a larger number of the particles. For a fixed repulsive Bose-Fermi interaction, the frequency shifts increase from the BCS side to the BEC regime, and the different speeds of the increases of frequency shifts, especially for the Fermi superfluid,
can be used to characterize different ground configurations of the mixtures \cite{bho2008,fla2007,sad2007,ufr2017} in different trap geometries.
This is because that the frequency shifts are not only originated from the Bose-Fermi interaction, but also sensitive to the spatial distributions of
the equilibrium Bose and Fermi densities. The Bose superfluid is localized in a small region in the trapping center
within the Fermi superfluid, which produces a depletion of the Fermi density due to the repulsive Bose-Fermi interaction.
As the interaction energy of the Fermi superfluid decreases from the BCS side to the BEC side, such depletion
becomes more pronounced and its boundary are steeper. Thus the overlap integrals of the spatial deviations of these two densities increase, which result in a stronger coupling and larger frequency shifts. In the BEC regime the depletion of the Fermi density in the center is completely,
which is accompanied by significant increases of the frequency shifts.
In recent experiments on the degenerate Bose-Fermi mixture of {$^{41}$}K-{$^6$}Li \cite{huang2018},
a rapid frequency upshift of the breathing mode of the bosons is observed, which is attributed by the emergent interface
when the mixture undergoes phase separation \cite{lous2018} by increasing the repulsive interspecies interaction.
We hope that our theoretical results can provide a reference for future
experiments on collective oscillation modes of Bose-Fermi superfluid mixtures in the BCS-BEC crossover.

This paper is organized as follows. The coupled hydrodynamic equations
are introduced in Sec.\;IIA, and within the scaling theory the coupled set of differential equations for
the relevant scaling parameters are derived in Sec.\;IIB. Subsequently, the dispersion relations of the
collective modes of the Bose-Fermi superfluid mixtures in cylindrically symmetric traps are obtained in Sec.\;IIC.
The explicit expressions for the overlap density integrals and the frequencies of the collective modes are presented in Sec.\;IID.
In Sec.\;III, the frequencies of the collective modes of the Bose-Fermi superfluid mixtures
in a realistic experimental setting and their physical properties for different trap geometries are discussed analytically, and the valid parameter regimes of the analytical approximations are demonstrated by the numerical calculations.
Sec.\;IV is for conclusion.

\section{Basic equations}

\subsection{Superfluid hydrodynamic model}

We consider a mixture of bosonic atoms and two spin components of fermionic atoms which is prepared in superfluid state at a low enough temperature.
The dynamic properties of the Bose-Fermi superfluid mixture can be described by coupled hydrodynamic equations for superfluid.
For bosons the hydrodynamic equations are given by \cite{dal1999,pet2002}
\bes\label{hdyb}
\bea & &\label{hydrobn}\frac{\partial n_b}{\partial t}+\nabla\cdot(n_b {\bf{v}}_b)=0, \\
& & \label{hydrobv}m_b\frac{\partial {{\bf v}_b}}{\partial t}+\nabla\Big[
V^b_{\rm ext}+g_{b}n_b+\frac{1}{2}m_b{\bf v}_b^2+g_{bf}n_f\Big]=0,
\eea
\ees
where Eq.(\ref{hydrobn}) is the equation of continuity for atomic density $n_b({\bf r},t)$ and the total number of
bosons is normalized by $N_b=\int n_b({\bf r},t) d{\bf r}$,
and Eq.(\ref{hydrobv}) for the velocity field ${\bf v}_b({\bf r}, t)$ establishes the irrotational and inviscid
nature of the superfluid motion. The trapping potential acting on bosons is cylindrically symmetric with the form
$V^b_{\rm ext}({\bf r})=m_b[\omega^2_{b\perp}(x^2+y^2)+\omega^2_{bz}z^2]/2$,
where $\omega_{b\perp}$ and $\omega_{bz}$ denotes the trapping frequencies and $m_b$ is the mass of a bosonic atom.
The boson-boson interaction strength is related to the s-wave scattering length $a_{b}$ by $g_{b}=4\pi\hbar^2a_b/m_b$.

The superfluid hydrodynamic equations for fermions in terms of the density $n_f({\bf r},t)$ and velocity field ${\bf v}_f({\bf r},t)$
are given by \cite{gio2008,lan1987}, respectively
\bes\label{hdyf}
\bea
& &\label{hydrofn}\frac{\partial n_f}{\partial t}+\nabla\cdot(n_f {\bf{v}}_f)=0, \\
& & \label{hydrofv}m_f\frac{\partial {{\bf v}_f}}{\partial t}+\nabla\Big[
V^f_{\rm ext}+\mu(n_f)+\frac{1}{2}m_f{\bf v}_f^2+g_{bf}n_b\Big]=0,
\eea
\ees
where the trapping potential acting on fermions is $V^f_{\rm ext}({\bf r})=m_f[\omega^2_{f\perp}(x^2+y^2)+\omega^2_{fz}z^2]/2$
with $m_f$ the mass of a fermionic atom, and the total number of fermions in superfluid state is normalized by $N_f=\int n_f({\bf r},t) d{\bf r}$.
In contrast to a simple expression of the interaction strength in the bosonic part, the two-spin fermionic
interaction is characterized by the equation of state $\mu(n_f)$. In order to obtain analytical results in various superfluid
regimes in a unified way, we take a polytropic approximation \cite{ma2005,wen2010}, i.e.
\bes\label{eosf}
\bea
\label{equf}& &\;\mu(n_f)=\mu_0(\frac{n_f}{n_0})^{\gamma}\;\;\;\;\;\;\;\;\mu_0=\epsilon_f[\sigma(\eta)-\frac{\eta}{5}\frac{\partial \sigma (\eta)}{ \partial \eta}],\\
\label{gamma}& &\gamma \equiv \gamma(\eta)=\frac{n_f}{\mu}\frac{\partial \mu}{\partial n_f}=
\frac{\frac{2}{3}\sigma (\eta)-\frac{2\eta}{5}\sigma^{\prime
}(\eta)+\frac{\eta^{2}}{15}\sigma^{\prime \prime}(\eta)}{\sigma (\eta)-\frac{\eta}{5}%
\sigma^{\prime }(\eta)},
\eea
\ees
where the reference chemical potential $\mu_0$ is proportional to the Fermi energy
$\epsilon_f=(\hbar k_f)^2/(2m_f)=\hbar(3N_f\omega^2_{f\perp}\omega_{fz})^{1/3}$ defined in a cylindrically symmetric trap
and reference atomic number density is given by the density of noninteracting Fermi gas at the trapping center
$n_0=(2m_f\epsilon_f)^{3/2}/(3\pi^2\hbar^3)$.
In order to be close to experimental observations, $\sigma (\eta)$ is based on the explicit expressions of fitting
functions from ENS experimental data \cite{nav2010}. The effective polytropic index $\gamma$ and reference chemical potential $\mu_0$
are determined by $\sigma (\eta)$ as a function of the dimensionless interaction $\eta=1/k_fa_f$, which have been plotted in Ref.\cite{wen2017}.
To describe the coupling between these two types of superfluid, we have introduced the boson-fermion interaction $g_{bf}=2\pi\hbar^2a_{bf}/m_{bf}$
at the mean-field level with boson-fermion scattering length $a_{bf}$ and reduced mass $m_{bf}=m_bm_f/(m_b+m_f)$.
It is worth noting that in the BEC limit where $1/k_fa_f\gg1$ and $a_{f}$ is comparable to $a_{bf}$,
the boson-fermion interaction should be replaced by the boson-dimer interaction \cite{ren2014,cui2014}.

For the coupled hydrodynamic equations (\ref{hdyb}) and (\ref{hdyf}), they actually work in the Thomas-Fermi (TF) regime, in which
they are analytically simpler to handle in the absence of the quantum pressure terms. The TF approximation is valid, provided that the interactomic interaction
energy is large enough to make the kinetic energy pressure negligible, i.e. in the large particle limit and collective excitations are of sufficiently
long wavelength. By including the proper quantum pressure terms \cite{lsal2008,cso2010}, the coupled hydrodynamic equations are equivalent to the coupled order-parameter
equations \cite{adki2008-2010,wen2018}.  In addition, the superfluid hydrodynamic equations only describe the dynamics of superfluid components, ignoring
single particle excitation, normal components and temperature effects.

\subsection{Scaling theory for a coupled system}

To account for collective oscillation modes in a coupled system, we resort to a scaling theory \cite{cas1996,kag1997}.
The basic idea behind the scaling method is to take appropriate scaling ansatz and simplify
time-dependent problems into solving differential equations for the scaling parameters. It is specially suited for
3D hydrodynamic equations, whereby numerical simulations are very expensive.
Moreover, it is enable to derive analytical approximations that provide a deep physical insight into the problem.
This technique was first proposed in the context of BEC \cite{cas1996,kag1997}, then used in the power-law
equation of state for superfluid Fermi gases in the BCS-BEC crossover \cite{men2002,hui2004,ast2005}. The
calculated collective mode frequencies are shown to be in quantitative agreement with experiments \cite{fermc,hui2004,ast2005}.
The extension of the scaling theory to a coupled system was developed in the cases
of degenerate Bose-Fermi mixtures \cite{liu2003,hui2003}.

The scaling anzatz for the time-dependent density profiles for the Bose and Fermi superfluids
are chosen as follows \cite{liu2003,hui2003}, respectively,
\bes\label{scalen}
\bea & &\label{scaleb}
n_b(x,y,z,t)=\frac{1}{b_x(t)b_y(t)b_z(t)}n^0_b(\frac{x}{b_x(t)},\frac{y}{b_y(t)},\frac{z}{b_z(t)}),\\
& & \label{scalef}
n_f(x,y,z,t)=\frac{1}{a_x(t)a_y(t)a_z(t)}n^0_f(\frac{x}{a_x(t)},\frac{y}{a_y(t)},\frac{z}{a_z(t)}),
\eea
\ees
where $n^0_b$ and $n^0_f$ are equilibrium density distributions for the Bose and Fermi superfluids, respectively.
The scaling anzatz for the velocity fields can be obtained by inserting the scaling anzatz Eqs.(\ref{scaleb}) and (\ref{scalef})
into the equation of continuity Eqs.(\ref{hydrobn}) and (\ref{hydrofn}), respectively
\bes\label{veloc}
\bea & &\label{velocb}
{\bf v}_b(x,y,z,t)=(\frac{x}{b_x}\frac{db_x}{dt},\frac{y}{b_y}\frac{db_y}{dt},\frac{z}{b_z}\frac{db_z}{dt}),\\
& & \label{velocf}
{\bf v}_f(x,y,z,t)=(\frac{x}{a_x}\frac{d a_x}{dt},\frac{y}{a_y}\frac{d a_y}{dt},\frac{z}{a_z}\frac{d a_z}{dt}).
\eea
\ees

Substituting the scaling ansatz for the densities (\ref{scalen}) and the velocity fields (\ref{veloc}) into the Eqs.(\ref{hydrobv}) and (\ref{hydrofv}),
we arrive at the differential equations for the scaling parameters
\bes\label{scalEq1}
\bea & &\label{scalEqb1}
R_{bi}\frac{d^2 b_{i}}{dt^2}+\omega^2_{bi}b_iR_{bi}+\frac{g_b}{m_bb_{i}\prod\limits_jb_j}\frac{\partial n^0_b({\bf R}_b)}{\partial R_{bi}}+\frac{g_{bf}}{m_bb_{i}\prod\limits_ja_j}\frac{\partial n^0_f({\bf R}_f)}{\partial R_{bi}}=0,\\
& & \label{scalEqf1}
R_{fi}\frac{d^2 a_{i}}{dt^2}+\omega^2_{fi}a_iR_{fi}+\frac{1}{m_fa_{i}}\frac{\partial \mu(n^0_f({\bf R}_f))}{\partial R_{fi}}+\frac{g_{bf}}{m_fa_{i}\prod\limits_jb_j}\frac{\partial n^0_b({\bf R}_b)}{\partial R_{fi}}=0,
\eea
\ees
where we have introduced the time-dependent coordinates ${\bf R}_b=[x/b_x(t), y/b_y(t), z/b_z(t)]$ and ${\bf R}_f=[x/a_x(t), y/a_y(t), z/a_z(t)]$.
It is seen that we transfer the time-dependent problems for the coupled hydrodynamic equations into
solving the ordinary differential equations for $b_i(t)$ and $a_i(t)$, with $i=x,y,z$ respectively, and the disturbations of
density profiles are expressed by these scaling parameters. In the equilibrium states Eqs.(\ref{scalEq1}) reduce to
\bes\label{scalEq2}
\bea & &\label{scalEqb2}
m_b\omega^2_{bi}r_i+g_b\frac{\partial n^0_b({\bf r})}{\partial r_i}+g_{bf}\frac{\partial n^0_f({\bf r})}{\partial r_{i}}=0,\\
& & \label{scalEqf2}
m_f\omega^2_{fi}r_i+\frac{\partial \mu(n^0_f({\bf r}))}{\partial r_{i}}+g_{bf}\frac{\partial n^0_b({\bf r})}{\partial r_{i}}=0.
\eea
\ees
In order to obtain scaling solutions for a coupled system, i.e. in the presence of the boson-fermion interation $g_{bf}$,
a useful strategy is developed that is assuming the scaling form of the solution a {\it priori} and fulfilling it on an average by integrating over
the spatial coordinates \cite{hui2003,liu2003}.  Combining the differential
equations (\ref{scalEq1}) with the equilibrium states (\ref{scalEq2}) and carrying out the spatial integration,
we obtain the following expressions
\bes\label{scalEq3}
\bea \label{scalEqb3}
\frac{d^2 b_{i}}{dt^2}+\omega^2_{bi}b_i&-&\frac{\omega^2_{bi}}{b_{i}\prod\limits_jb_j}+\frac{g_{bf}}{N_b m_b\langle R^2_i\rangle}_b\frac{1}{ b_i\prod\limits_jb_j}
\int d{\bf R}_b\;R_{bi}n^0_b(\frac{\bf a}{\bf b}{\bf R}_b)
\frac{\partial}{\partial R_{bi}} n^0_f({\bf R_b})\\\nonumber
& &-\frac{g_{bf}}{N_b m_b\langle R^2_{i}\rangle_b}\frac{1}{b_i\prod\limits_jb_j}
\int d{\bf R}_b\;R_{bi}n^0_b({\bf R}_b)\frac{\partial}{\partial R_{bi}} n^0_f({\bf R}_b) =0,\\
\label{scalEqf3}
\frac{d^2 a_{i}}{dt^2}+\omega^2_{fi}a_i&-&\frac{\omega^2_{fi}}{a_{i}(\prod\limits_ja_j)^{\gamma}}+\frac{g_{bf}}{N_fm_f\langle R^2_i\rangle_f}\frac{1}{a_i\prod\limits_ja_j}
\int d{\bf R}_f\;R_{fi}n^0_f(\frac{{\bf b}}{{\bf a}}{\bf R}_f)
\frac{\partial}{\partial R_{fi}} n^0_b({\bf R}_f)\\\nonumber
& &-\frac{g_{bf}}{N_f m_f\langle R^2_i\rangle_fa_i(\prod\limits_ja_j)^{\gamma}}
\int d{\bf R}_f\;R_{fi}n^0_f({\bf R}_f)\frac{\partial}{\partial R_{fi}}n^0_b({\bf R}_f) =0,
\eea
\ees
with ${\bf b}\equiv[b_x,b_y,b_z]$ and ${\bf a}\equiv[a_x,a_y,a_z]$.
$\langle R^2_{i} \rangle_b=(1/N_b)\int d{\bf R}_b n^0_b({\bf R}_b)R^2_{bi}$
and $\langle R^2_{i} \rangle_f= (1/N_f)\int d{\bf R}_f n^0_f({\bf R}_f)R^2_{fi}$
correspond to the mean square radii of the Bose and Fermi superfluids in the $i$ axis, respectively.

Due to the collective oscillations considered
here are small around the equilibrium states, by expanding
$n^0_b(\frac{{\bf a}}{\bf b}{\bf R}_b)\simeq n^0_b({\bf R}_b)+\sum\limits_{k}\frac{\partial n^0_b({\bf R}_b)}{\partial R_{bk}}(\frac{a_k}{b_k}-1)R_{bk}$ and $n^0_f(\frac{{\bf b}}{{\bf a}}{\bf R}_f)\simeq n^0_f({\bf R}_f)
+\sum\limits_{k}\frac{\partial n^0({\bf R}_f)}{\partial R_{fk}}(\frac{b_k}{a_k}-1)R_{kf}$ one can simplify
Eqs.(\ref{scalEq3}) as
\bes\label{scalEq4}
\bea \label{scalEqb4}
& &\frac{d^2b_i}{dt^2}+\omega^2_{bi}b_i-\frac{\omega^2_{bi}}{b_i\prod\limits_jb_j}+\sum\limits_k\frac{\omega^2_{bi}}{b_i\prod\limits_jb_j}
(\frac{a_k}{b_k}-1)B_{ik}=0,\\
\label{scalEqf4}
& &\frac{d^2a_i}{dt^2}+\omega^2_{fi}a_i-\frac{\omega^2_{fi}}{a_i(\prod\limits_ja_j)^{\gamma}}+\frac{\omega^2_{fi}}{a_i}(\frac{1}{\prod\limits_j a_j}-\frac{1}{(\prod\limits_ja_j)^{\gamma}})F_i+\sum\limits_k\frac{\omega^2_{fi}}{a_i\prod\limits_ja_j}(\frac{b_k}{a_k}-1)F_{ik}=0.
\eea
\ees
The dimensionless parameters proportional to $g_{bf}$ are given by
\bes\label{coupledin}
\bea & &\label{coupledin1}
B_{ik}=\frac{g_{bf}}{N_bm_b\omega^2_{bi}\langle R^2_{i}\rangle}_b\int d{\bf r}\; \frac{\partial n^0_f({\bf r})}{\partial r_i}r_ir_k\frac{\partial n^0_b({\bf r})}{\partial r_k} \\
& & \label{coupledin2}
F_i=\frac{g_{bf}}{N_fm_f\omega^2_{fi}\langle R^2_{i}\rangle}_f\int d{\bf r}\;\frac{\partial n^0_b({\bf r})}{\partial r_i} r_i  n^0_f({\bf r})\\
& & \label{coupledin3}
F_{ik}=\frac{g_{bf}}{N_fm_f\omega^2_{fi}\langle R^2_{i}\rangle}_f\int d{\bf r}\;\frac{\partial n^0_b({\bf r})}{\partial r_i}r_ir_k \frac{\partial n^0_f({\bf r})}{\partial r_k}
,
\eea
\ees
where we have replaced the variables ${\bf R}_b$ and ${\bf R}_f$ by ${\bf r}$ for simplicity in Eqs.(\ref{coupledin}),
because the each integration is relevant to either ${\bf R}_b$ or ${\bf R}_f$.

\subsection{Collective oscillation modes}

\begin{figure}
\includegraphics[scale=0.25]{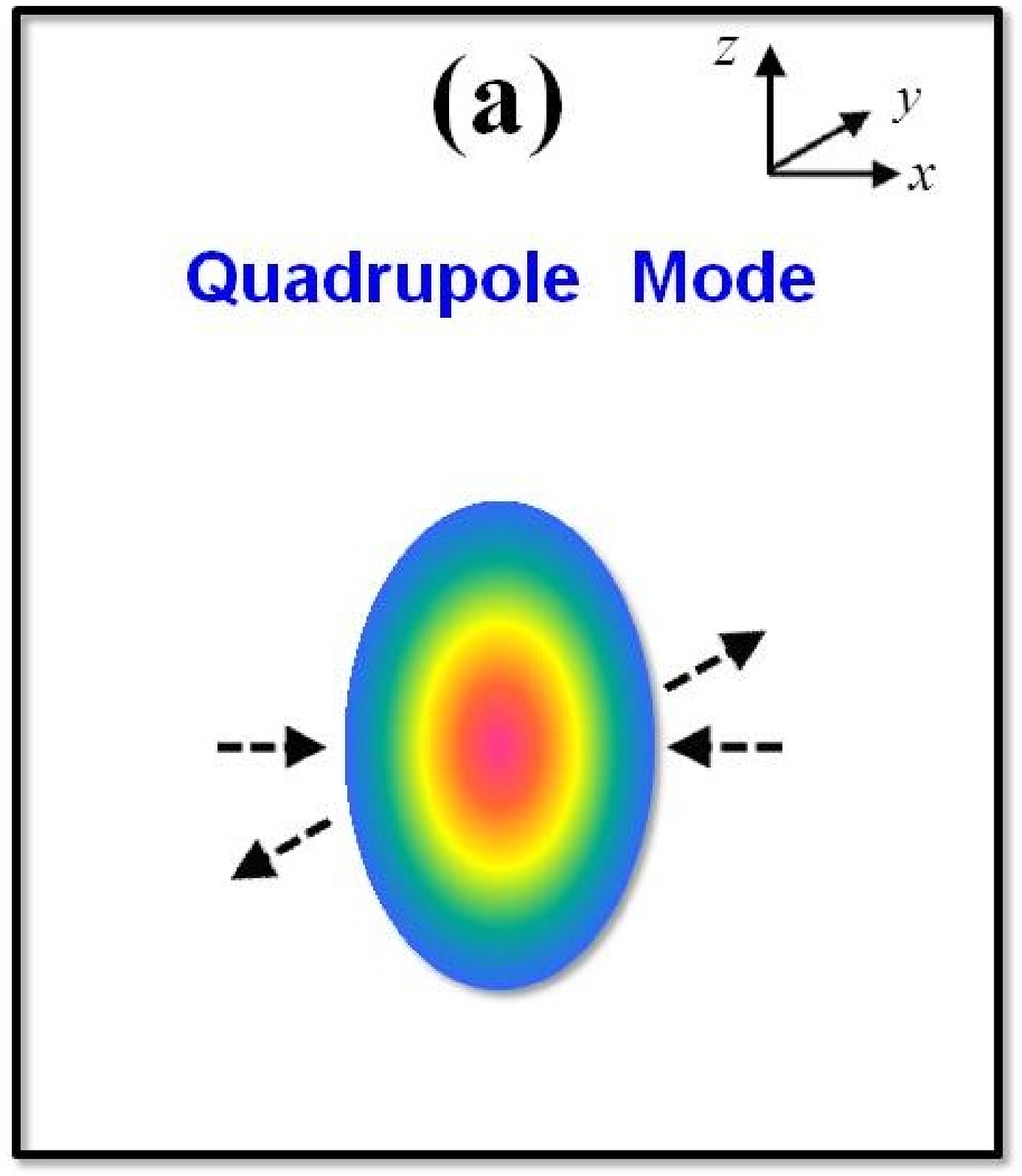}
\includegraphics[scale=0.25]{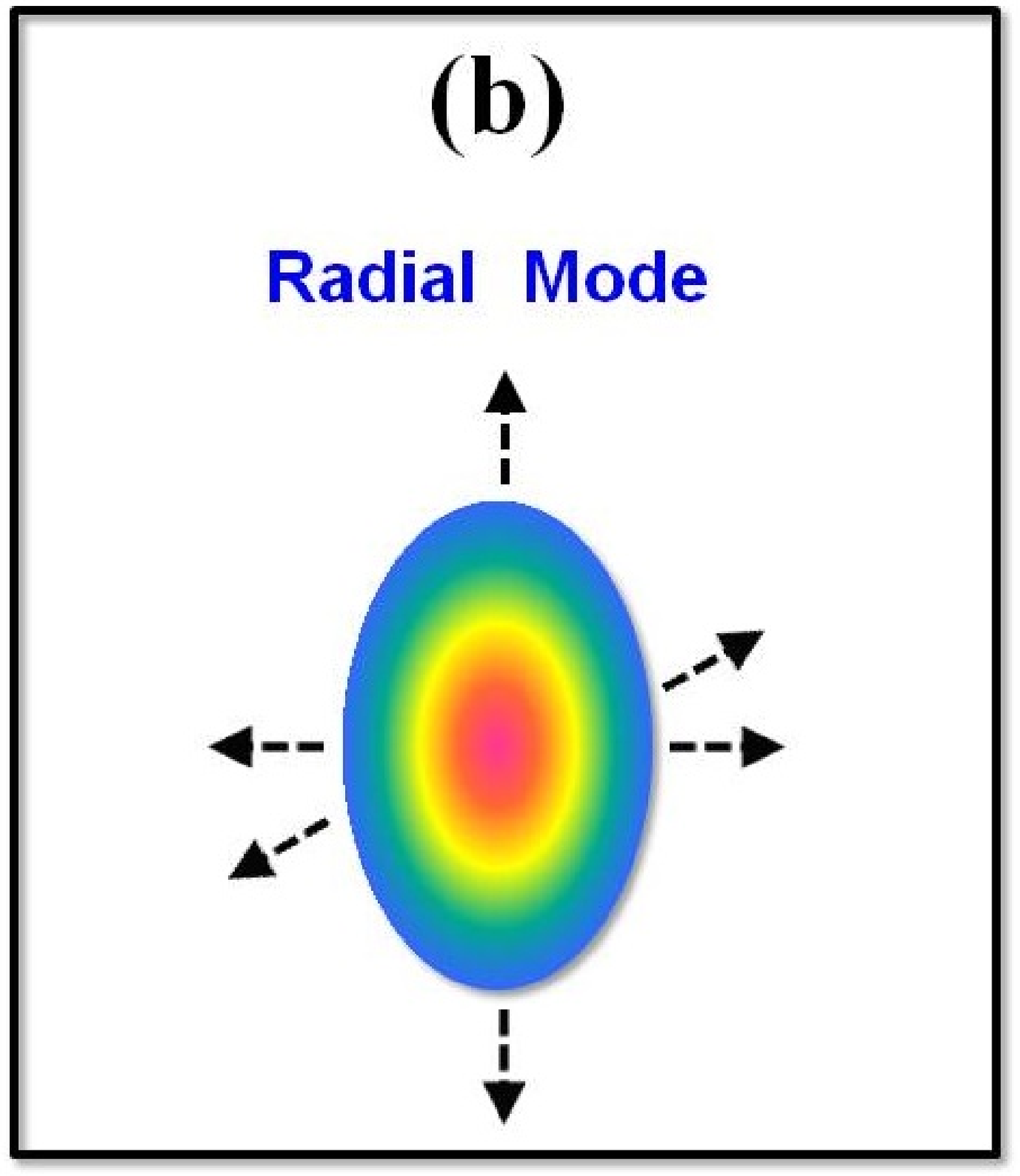}
\includegraphics[scale=0.25]{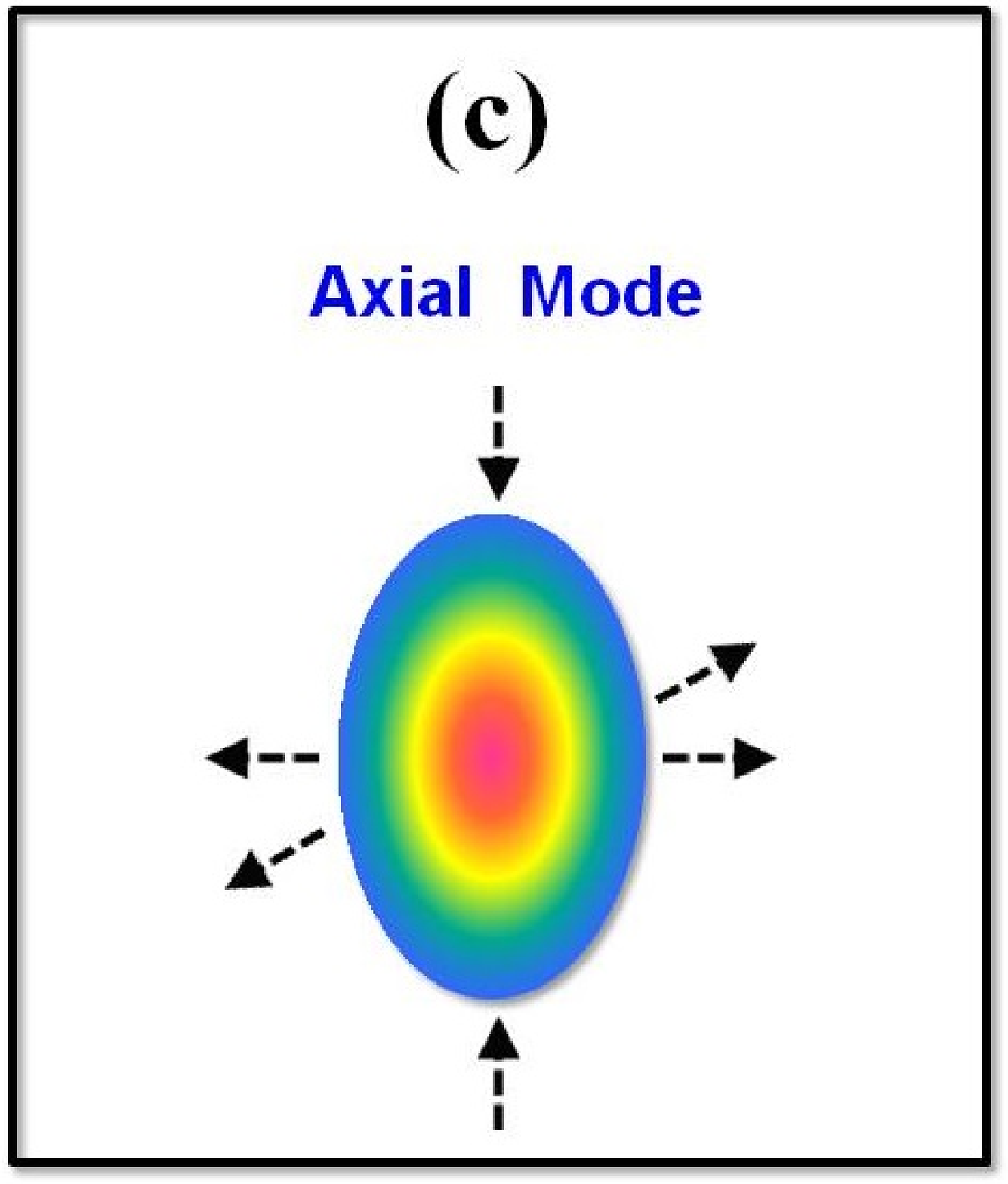}
\caption{\footnotesize{A schematic illustration for collective modes of superfluids in a cylindrically symmetric trap
obtained from the scaling method: (a) (radial) quadrupole  mode, (b) radial mode, and (c) axial mode. }}
\end{figure}

In order to clearly understand the collective oscillation modes obtained from the scaling method, let us
first discuss the results without the boson-fermion interaction $g_{bf}$. In this case
all dimensionless parameters (\ref{coupledin}) disappear, and Eqs.(\ref{scalEqb4}) and (\ref{scalEqf4}) decouple and
describe the Bose and Fermi superfluids alone, respectively.
In a cylindrical symmetry, the frequencies of the harmonic traps are $\omega_{bx}(\omega_{fx})=\omega_{by}(\omega_{fy})=\omega_{b\perp}(\omega_{f\perp}$)
and $\omega_{bz}(\omega_{fz})=\lambda_b\omega_{b\perp}(\lambda_f\omega_{f\perp})$, with the trap anisotropy $\lambda_b=\omega_{bz}/\omega_{b\perp}$ ($\lambda_f=\omega_{fz}/\omega_{f\perp}$)
for the Bose (Fermi) superfluids. The resulting eigenvalues of the scaling equations
are three mode frequencies \cite{vic1996,hei2004}. By examining the signs of the eigenvectors, one can find that one is (radial)
quadrupole mode illustrated in Fig.1(a), which only supports in the radial plane with no excitation in axial direction, and the radii oscillating out of phase with each other. The frequency of the quadrupole mode \cite{sstri1996,vic1996,adhk2010}
is, respectively, for Bose and Fermi superfluids
\be\label{mr}
\omega^2_{bq}/\omega^2_{b\perp}=2,\;\;\;\;\;\;\;\;\;\;\;\;\;\;\;\;\; \omega^2_{fq}/\omega^2_{f\perp}=2.
\ee
The other two mode frequencies for Bose and Fermi superfluids are given by \cite{vic1996,hui2004,ast2005}
\bes\label{mm}
\bea & &\label{bmm}
\omega^2_{b\pm}/\omega^2_{b\perp}=2+\frac{3}{2}\lambda^2_b\pm\frac{1}{2}
\sqrt{9\lambda_{b}^4-16\lambda^2_{b}+16},\\
& & \label{fmm}
\omega^2_{f\pm}/\omega^2_{f\perp}=(\gamma+1)+\frac{\gamma+2}{2}\lambda^2_f
\pm\sqrt{[1+\gamma+\frac{\gamma+2}{2}\lambda^2_{f}]^2-2(2+3\gamma)\lambda^2_{f}},
\eea
\ees
where $\pm$ refer to the radial and axial mode, respectively. As shown in Fig.\;1(b) and Fig.\;1(c), respectively,
the radial mode features an in-phase oscillation for all radii in the radial and
axial directions, while the axial mode corresponds to two radii in radial direction oscillating in phase with each other,
but out of phase with the radius in the axial direction. Differently from the quadrupole mode, the radial and axial modes are relevant to the anisotropy of the trap. In a highly elongated trap ($\lambda_{b,f}\ll1$)
\cite{hui2004,ast2005,fermc}, the frequency of the radial mode
reduces to $\omega_{b+}=2\omega_{b\perp}(\omega_{f+}=\sqrt{2(\gamma+1)}\omega_{f\perp})$, which coincides with the radial
(transverse) breathing mode \cite{vic1996,hei2004,huang2018}, and the axial mode is $\omega_{b-}=\sqrt{5/2}\omega_{bz}(\omega_{f-}=\sqrt{(3\gamma+2)/(\gamma+1)}\omega_{fz})$.
This is the reason why they are called radial and axial mode, respectively. In the oblate limit ($\lambda_{b,f}\gg1$), the frequency of radial mode reduces to $\omega_{b+}=\sqrt{3}\omega_{bz}(\omega_{f+}=\sqrt{\gamma+2}\omega_{fz})$, and the axial mode is $\omega_{b-}=\sqrt{10/3}\omega_{b\perp}(\omega_{f-}=\sqrt{(6\gamma+4)/(\gamma+2)}\omega_{f\perp})$.
In a spherical trap ($\lambda_{b,f}=1$), the frequency of the radial mode  is $\omega_{b+}=\sqrt{5}\omega_b (\omega_{f+}=\sqrt{3\gamma+2}\omega_f)$, which is also referred as monopole mode \cite{sstri1996}, and the axial mode frequency is $\omega_{b-}=\sqrt{2}\omega_b(\omega_{f-}=\sqrt{2}\omega_f)$, recovering to the quadrupole mode Eqs.(\ref{mr}).

In the presence of the Bose-Fermi interaction $g_{bf}$, the collective modes of the Bose and Fermi superfluids are coupled each other
and their frequencies are varied. For the quadrupole mode, the linearation of Eqs.(\ref{scalEq4}) around the equilibrium states result in the eigenvalue function
$ d^2 {{\bf P}}/dt^2= M_q{\bf P}$, where the vector notation is ${\bf P}^{T}\equiv (\delta b_{x}, \delta b_{y}, \delta a_{x}, \delta a_{y})$ and matrix $M_q$
is written in a cylindrical coordinate ($\bot, z$) by
\begin{equation}\label{matrixr}
M_q=
\left(
\begin{array}{cccc}
(3-\frac{3}{4}B_{\rho\rho})\omega^2_{b\perp}&(1-\frac{1}{4}B_{\rho \rho})\omega^2_{b\perp}&\;\;\frac{3}{4}B_{\rho\rho}\omega^2_{b\perp}&\;\;\frac{1}{4}B_{\rho \rho}\omega^2_{b\perp}\\
(1-\frac{1}{4}B_{\rho\rho})\omega^2_{b\perp}&(3-\frac{3}{4}B_{\rho\rho})\omega^2_{b\perp}&\frac{1}{4}B_{\rho\rho}\omega^2_{b\perp}&\frac{3}{4}B_{\rho\rho}\omega^2_{b\perp}\\
\frac{3}{4}F_{\rho\rho}\omega^2_{f\perp}&\frac{1}{4}F_{\rho \rho}\omega^2_{f\perp}&G\omega^2_{f\perp}&H\omega^2_{f\perp}\\
\frac{1}{4}F_{\rho\rho}\omega^2_{f\perp}&\frac{3}{4}F_{\rho\rho}\omega^2_{f\perp}&H\omega^2_{f\perp}&G\omega^2_{f\perp}\\
\end{array}
\right)
\end{equation}
with $G=2+\gamma+(\gamma-1)F_{\rho}-3F_{\rho\rho}/4$ and $H=\gamma+(\gamma-1)F_{\rho}-F_{\rho \rho}/4$.
The dimensionless parameters $B_{\rho\rho}$, $F_{\rho\rho}$, and $F_\rho$, which
are integrals in terms of the density profiles of the Bose and Fermi superfluids at equilibrium, are found to be
responsible for the coupling of the Bose and Fermi components.

For the radial and axial mode, the eigenvector corresponds to ${\bf P}^{T}=(\delta b_{\perp}, \delta b_{z}, \delta a_{\perp},\delta a_{z})$
and the matrix $M$ is defined by
\begin{equation}\label{matrixm}
M=\left(
\begin{array}{cccc}
(4-B_{\rho\rho})\omega^2_{b\perp}&(1-B_{\rho z})\omega^2_{b\perp}&\;\;B_{\rho\rho}\omega^2_{b\perp}&\;\;B_{\rho z}\omega^2_{b\perp}\\
(2-B_{z\rho})\omega^2_{bz}&(3-B_{zz})\omega^2_{bz}&B_{z\rho}\omega^2_{bz}&B_{zz}\omega^2_{bz}\\
F_{\rho\rho}\omega^2_{f\perp}&F_{\rho z}\omega^2_{f\perp}&M_{\rho\rho}\omega^2_{f\perp}&M_{\rho z}\omega^2_{f\perp}\\
F_{z\rho}\omega^2_{fz}&F_{zz}\omega^2_{fz}&M_{z\rho}\omega^2_{fz}&M_{zz}\omega^2_{fz}\\
\end{array}
\right)
\end{equation}
with $M_{\rho\rho}=2\gamma+2+2(\gamma-1)F_{\rho}-F_{\rho\rho}$, $M_{\rho z}=\gamma+(\gamma-1)F_{\rho}-F_{\rho z}$,
$M_{z \rho}=2\gamma+2(\gamma-1)F_z-F_{z \rho}$ and $M_{z z}=\gamma+2+(\gamma-1)F_{z}-F_{z z}$. The expressions of the dimensionless parameters proportional to $g_{bf}$
in the cylindrical coordinates ($\alpha,\beta=\rho(\perp),z$) take the forms
\bes\label{dimenpara}
\bea
B_{\alpha\beta}&=&\frac{g_{bf}}{N_bm_b\omega^2_{b\alpha}{\langle R^2_{\alpha}\rangle}_b}\int d{\bf r}\; \frac{\partial n^0_f}{\partial r_\alpha}r_\alpha r_\beta\frac{\partial n^0_b}{\partial r_\beta}\\
F_{\alpha\beta}&=&\frac{g_{bf}}{N_fm_f\omega^2_{f\alpha}\langle R^2_{\alpha}\rangle_f} \int  d{\bf r} \;\frac{\partial n^0_b}{\partial r_\alpha} r_\alpha r_\beta\frac{\partial n^0_f}{\partial r_\beta}\\
F_{\alpha}&=&\frac{g_{bf}}{N_fm_f\omega^2_{f\alpha}{\langle R^2_{\alpha}\rangle}_f}\int d{\bf r}\; \frac{\partial n^0_b}{\partial r_\alpha} r_\alpha  n^0_f,
\eea
\ees
with the mean square radii in the $\alpha=\perp, z$ direction given by $\langle R_\alpha^2\rangle_b= (1/N_b)\int d{\bf r}\; r_\alpha^2 n^0_b$ for the Bose superfluid
and $\langle R_\alpha^2 \rangle_f= (1/N_f)\int d{\bf r} \; r_\alpha^2n^0_f $ for the Fermi superfluid.

\subsection{Analytical approximations for the integral terms}

In the previous subsection, one can find that the dimensionless parameters are the spatial overlap integrals in
terms of the equilibrium density profiles of the Bose and Fermi superfluids, and the frequencies
of collective oscillation modes of Bose-Fermi superfluid mixtures crucially depend on these integrals.
Eqs.\;(\ref{hydrobv}) and (\ref{hydrofv}) at groundstates give the density profiles
of the Bose and Fermi superfluids coupled each other, which are written in cylindrical coordinate as
\bes\label{eqden}
\bea & &\label{bos}
n^0_b(r,z)=\frac{1}{g_b}\left[\mu_b-\frac{1}{2}m_b(\omega^2_{b\perp}r^2+\omega^2_{b z}z^2)-g_{bf}n^0_f(r,z)\right],\\
& & \label{fer}
n^0_f(r,z)=\frac{n_0}{\mu^{1/\gamma}_0}\left[\mu_f-\frac{1}{2}m_f(\omega^2_{f \perp}r^2+\omega^2_{f z}z^2)-g_{bf}n^0_b(r,z)\right]^{1/\gamma},
\eea
\ees
where the bulk chemical potentials $\mu_b$ and $\mu_f$ are determined by the total numbers of bosons and fermions,
respectively. Without the boson-fermion interaction $g_{bf}$, the explicit expressions for
the density profile of the Bose superfluid are \cite{dal1999,pet2002}
\be\label{Bprofile0}
     n^{00}_{b}({r,z})=\frac{1}{g_{b}}{\rm max}\big[\mu_{b}-V^b_{\rm ext}(r,z),0\big],
     \;\;\;\;\;\; \mu_{b}=(\frac{15N_b\hbar^2a_b\sqrt{m_b}\omega^2_{b\perp}\omega_{bz}}{4\sqrt{2}})^{2/5},\\
\ee
and the Fermi density profiles along the BCS-BEC crossover \cite{wen2010} are
\begin{eqnarray}\label{Fprofile0}
& & n^{00}_{f}(r,z)=\frac{n_0}{\mu^{1/\gamma}_0}{\rm max}\big[\mu_{f}-V^f_{\rm ext}(r,z),0\big]^{1/\gamma},\\\nonumber
&&\mu_{f}=\epsilon_f\Big[(\sigma(\eta)-\frac{\eta\sigma^{\prime}(\eta)}{5})^{\frac{1}{\gamma}}\frac{\pi^{\frac{1}{2}}
\Gamma({\frac{1}{\gamma}+\frac{5}{2}})}{8\Gamma({\frac{1}{\gamma}+1})}\Big]^{2\gamma/(2+3\gamma)}.
\end{eqnarray}
The density profiles $n^{00}_b$ and $n^{00}_f$ in
the absence of $g_{bf}$ can be regarded as the zero-order approximation for the density profiles (\ref{bos}) and (\ref{fer}). By
a perturbative expansion, the density profiles $n^{01}_b$ and $n^{01}_f$  at the first-order approximation can be
naturally obtained by replacing $n^0_f$ by the zero-order results $n^{00}_f$ in Eq.(\ref{bos})
and $n^0_b$ by $n^{00}_b$ in Eq.(\ref{fer}), correspondingly. Differently from the previous works by the numerical methods \cite{hui2003,liu2003},
we apply analytical approximations for the integrals \cite{wen2017}
that holds in the recent experimental situations, then give explicit expressions for the dimensionless parameters.

As a example of the integral $I_{\rho\rho}=\int \rho d\rho dz\frac{\partial n^{0}_b}{\partial \rho}\rho^2 \frac{\partial n^{0}_f}{\partial \rho}$,
we derive it as following
\bes\label{interr}
\bea
& & \label{interr1} I_{\rho\rho}=\int \rho d\rho dz \frac{\rho}{g_b}\left[-m_b\omega^2_{b\perp}+g_{bf}\frac{n_0}{\gamma\mu^{1/\gamma}_0}(\mu_f-V_{\rm ext}^f)^{1/\gamma-1}m_f\omega^2_{f\perp}\right]\rho^2\\\nonumber
& & \;\;\;\;\;\;\;\rho\frac{n_0}{\gamma\mu_0^{1/\gamma}}\left[\mu_f-V^f_{\rm ext}-g_{bf}n_b\right]^{1/\gamma-1}(-m_f\omega^2_{f\perp}+\frac{g_{bf}}{g_b}m_b\omega^2_{b\perp}),\\
& & \label{interr2} \approx\frac{m^2_b\omega^4_{b\perp}}{g_b}\left[g_{bf}(\frac{\partial n^{00}_f}{\partial \mu_f})_{{\bf r}=0}\frac{m_f\omega^2_{f\perp}}{m_b\omega^2_{b\perp}}-1\right]
(\frac{\partial n^{00}_f}{\partial \mu_f})_{{\bf r}=0}(\frac{g_{bf}}{g_b}-\frac{m_f\omega^2_{f\perp}}{m_b\omega^2_{b\perp}})\int_{V_B}d\rho dz \rho^5,\\
& & \label{interr3} =\frac{8}{7\pi}\mu_bN_b(\frac{\partial n^{00}_f}{\partial \mu_f})|_{{\bf r}=0}\left[g_{bf}(\frac{\partial n^{00}_f}{\partial \mu_f})|_{{\bf r}=0}
\frac{m_f\omega^2_{f\perp}}{m_b\omega^2_{b\perp}}-1\right]\left[\frac{g_{bf}}{g_b}-\frac{m_f\omega^2_{f\perp}}{m_b\omega^2_{b\perp}}\right],
\eea
\ees
where $(\frac{\partial n^{00}_f}{\partial \mu_f})_{{\bf r}=0}=\frac{n_0\mu_f^{1/\gamma-1}}{\gamma\mu^{1/\gamma}_0}$.
The recent experimental conditions for the Bose-Fermi superfluid mixtures \cite{fer2014,roy2017,yuping2018}
mainly share two common characteristics. First, the numbers of bosons and fermions are both large
at order of $10^4-10^6$ magnitude, and the number of bosons is one order smaller than fermions, thus the bosons
can be regarded as a mesoscopic impurity immersed in Fermi superfluids. Second, the boson-fermion interaction $g_{bf}$ is small.
Since $\mu_f$($N_f$) is much larger than $\mu_b$($N_b$) and $n_f$ is much smaller than $n_b$ due to the stronger interaction,
the last terms on the right sides of (\ref{bos}) and (\ref{fer}) are relatively smaller than the first terms and the Bose and Fermi density profiles are weakly coupled.

By using a perturbative method, in the first step of the integration (\ref{interr1}), we have substituted the density profiles
$n^0_b$ and $n^0_f$ by the first-order approximations $n^{01}_b$ and $n^{01}_f$, respectively. Since the Bose superfluid
is weakly interacting and the atomic number is smaller than the fermionic counterpart, the spatial distribution of the Bose superfluid
only overlaps with the Fermi superfluid in a small central region. In the second step (\ref{interr2}),
we approximate the Fermi density by the central value and the region of integration is the volume $V_B$ of the Bose
superfluid. Based on the above analysis for the parameters and these approximations, the explicit expression for the integral $I_{\rho\rho}$ is presented in
(\ref{interr3}). The same analysis can be also applied at the zero-order approximation, i.e. the density profiles in the integrals are replaced by $n^{00}_b$ and $n^{00}_f$,
and the integral is given by $I_{\rho\rho}=\frac{8}{7\pi}
N_b\mu_b{(\frac{\partial n^{00}_f}{\partial \mu_f}})_{{\bf r}=0}(\frac{m_f\omega^2_{f\perp}}{m_b\omega^2_{b\perp}})$. Compared with
the first-order approximation (\ref{interr3}), the result of the zero-order approximation is lack of the last term which is relevant to the ratio of boson-fermion
interaction $g_{bf}$ to boson interaction $g_b$. In a recent work \cite{wen2017}, we use the scaling theory to study the dipole mode
of the Bose-Fermi superfluid mixture, and the result for the frequency shift at the zero-order approximation reproduces
the mean-field model \cite{fer2014}. The explicit expressions for all involved integrals and the dimensionless parameters are presented in Appendix.

By examining the dispersion relations (\ref{matrixr}) and (\ref{matrixm}), one can find that if the elements for the couplings
of the amplitudes of the bosonic and fermionic excitations in the matrix are small which implies that
the effective coupling is weak, the explicit expressions for the frequencies of the collective oscillation modes
can be presented \cite{liu2003}. For the quadrupole mode, the high-(+)  and low-lying(-)
frequencies are explicitly expressed by
\bes\label{nmr}
\bea
{{\omega}^{\pm}_{q}}^2&=&\frac{1}{2}\Big\{(2-\frac{1}{2}B_{\rho\rho})\omega^2_{b\perp}+(2-\frac{1}{2}D_{\rho\rho})\omega^2_{f\perp} \\ \nonumber
 \;\;\;\;\;\;\;\;\;&\pm &\sqrt{\big[(2-\frac{1}{2}B_{\rho\rho})\omega^2_{b\perp}-(2-\frac{1}{2}D_{{\rho}{\rho}})\omega^2_{f\perp}\big]^2+B_{\rho\rho}D_{\rho\rho}\omega^2_{b\perp}\omega^2_{f\perp}}\Big\}.
\eea
\ees
The expressions for the high- and low-lying frequencies for the radial $\omega^{\pm}_{+}$ and axial
$\omega^{\pm}_{-}$ mode are more complicated and given by, respectively
\bes\label{nmm}
\bea
\label{nmmh} {\omega^{\pm}_{+}}^2&=&\frac{1}{2(M^{+}_b+1)(M^{+}_f+1)}\Big\{A^+_{22}(M^+_b+1)+A^+_{11}(M^+_f+1)\\\nonumber
\;\;\;\;\;&\pm &\sqrt{[A^+_{22}(M^{+}_b+1)-A^+_{11}(M^{+}_f+1)]^2+4(M^+_b+1)(M^+_f+1)A^+_{12}A^+_{21}}\Big\},\\
\label{nmml} {\omega^{\pm}_{-}}^2&=&\frac{1}{2(M^{-}_b+1)(M^{-}_f+1)}\Big\{A^-_{22}(M^-_b+1)+A^-_{11}(M^-_f+1)\\\nonumber
\;\;\;\;\;&\pm &\sqrt{[A^-_{22}(M^{-}_b+1)-A^-_{11}(M^{-}_f+1)]^2+4(M^-_b+1)(M^-_f+1)A^-_{12}A^-_{21}}\Big\},
\eea
\ees
with
\bes\label{nmm1}
\bea
& & \label{nmm11} A^\pm_{11}=\big[M^\pm_b(4-B_{\rho\rho})+(1-B_{\rho z})\big]\omega^2_{b\perp}+[M^\pm_b(2-B_{z\rho})+(3-B_{zz})]\omega^2_{bz}, \\
& & \label{nmm12} A^\pm_{12}=(M^\pm_fB_{\rho\rho}+B_{\rho z})\omega^2_{b\perp}+(M^\pm_fB_{z\rho}+B_{zz})\omega^2_{bz},\\
& & \label{nmm21} A^\pm_{21}=(M^\pm_bF_{\rho\rho}+F_{\rho z})\omega^2_{f\perp}+(M^\pm_bF_{z\rho}+F_{zz})\omega^2_{fz},\\
& & \label{nmm22} A^\pm_{22}=(M^\pm_fM_{\rho\rho}+M_{\rho z})\omega^2_{f\perp}+(M^\pm_fM_{z\rho}+M_{zz})\omega^2_{fz},
\eea
\ees
where the ratios of radial to axial excitation amplitude are given by $M^{\pm}_b=\omega^2_{b\perp}/(\omega^2_{b\pm}-4\omega^2_{b\perp})$ and $M^{\pm}_f=\gamma\omega^2_{f\perp}/[\omega^2_{f{\pm}}-(2\gamma+2)\omega^2_{f\perp}]$,
with $\omega_{b\pm}$ and $\omega_{f\pm}$ being the frequencies of the radial (+) and axial (-)
modes of the Bose and Fermi superfluids alone in Eqs.\;(\ref{mm}).
The positive (+) and negative (-) roots of the quadrupole ${\omega}^{\pm}_q$, radial $\omega^{\pm}_{+}$,
and axial $\omega^{\pm}_{-}$ mode frequencies are the relevant mode frequencies
for the Bose and Fermi superfluids, respectively.

\section{results and discussions}

In this section, we apply the theoretical results of the previous section to a realistic experimental situation as an example to discuss how the
Bose-Fermi interaction affects the frequencies of the collective modes of the Bose and Fermi superfluids in the BCS-BEC crossover
in different trap geometries. Our theoretical results can be also easily extended to other experimental settings. We choose the parameters of the
experiment performed at University of Science and Technology of China (USTC) on the dipole oscillations of a Bose-Fermi superfluid mixture of {$^{41}$}K-{$^6$}Li with a large mass-imbalance \cite{yuping2018}.
The experiment is performed in a cigar-shaped trap  ($\lambda_b=\omega_{bz}/\omega_{b\perp}=0.04, \lambda_f=\omega_{fz}/\omega_{f\perp}=0.06$),
with the radial and axial frequencies of the harmonic trap for bosons (fermions) being
$\omega_{b \perp}=2\pi\times 170.7$ $\rm{Hz}$ ($\omega_{f \perp}=2\pi\times 295.4$ $\rm{Hz}$)
and $\omega_{bz}=2\pi\times 6.295$ $\rm{Hz}$ ($\omega_{fz}=2\pi\times 16.453$ $\rm{Hz}$).
In the following calculations, the frequency in the radial direction is fixed, and the
axial frequency and the anisotropy $\lambda_b=\omega_{bz}/\omega_{b\perp}$($\lambda_f=\omega_{fz}/\omega_{f\perp}$) are varied
to realize different trap geometries.
For a spherical trap ($\lambda_b=1,\lambda_f=1$), the frequencies are given by
$\omega_{b}=2\pi\times 170.7$ $\rm{Hz}$ and $\omega_{f}=2\pi\times 295.4$ $\rm{Hz}$.
For a disk-shaped trap ($\lambda_b=7, \lambda_f=11$), the axial frequencies are enlarged by a same factor
$\omega_{bz}=2\pi\times 6.295\times189=2\pi\times 1190$ $\rm{Hz}$ and $\omega_{fz}=2\pi\times 16.453\times189=2\pi\times3109$ $\rm{Hz}$.
The total numbers of $^{41}$K bosons and $^{6}$Li fermions are given by $N_b=2.3\times10^5$ and
$N_f=1\times10^6$, respectively. The scattering lengths of boson-boson $a_b=60.5a_0$ and boson-fermion $a_{bf}=60.2a_0$
($a_0$ the Bohr radius) are fixed. The scattering length $a_f$ of two-spin fermionic atoms is tunable
across the BCS-BEC crossover through a Feshbach resonance.
\begin{figure}
\includegraphics[scale=0.3]{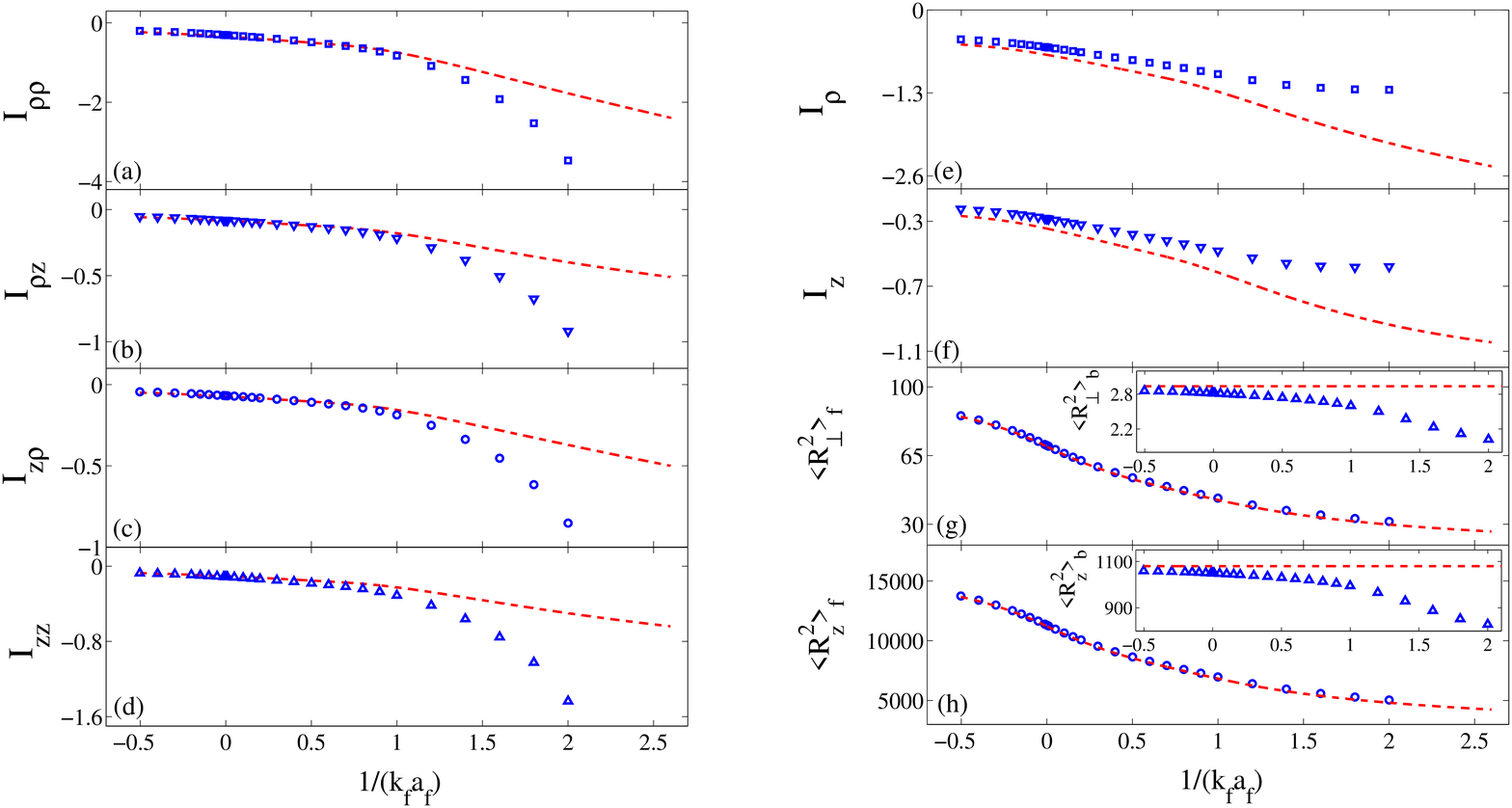}
\caption{\footnotesize{The analytical and numerical results for the respective integrals in units of
$10^6a^{-3}_{\rm ho}$\;($a_{\rm ho}\equiv\sqrt{\hbar/(m_b\omega_{b\perp})}$) evaluated with
the equilibrium Bose and Fermi superfluid densities: (a) $I_{\rho\rho}=\int \rho d\rho dz
\frac{\partial n^0_b}{\partial \rho}\rho^2\frac{\partial n^0_f}{\partial \rho}$, (b) $I_{\rho z}=\int \rho d\rho dz
\frac{\partial n^0_f}{\partial \rho}z\rho\frac{\partial n^0_b}{\partial z}$, (c) $I_{z \rho}=\int \rho d\rho dz
\frac{\partial n^0_f}{\partial z}z\rho\frac{\partial n^0_b}{\partial \rho}$, (d) $I_{z z}=\int \rho d\rho dz
\frac{\partial n^0_b}{\partial z}z^2\frac{\partial n^0_f}{\partial z}$, and (e) $I_{\rho}=\int \rho d\rho dz
n^0_f\rho\frac{\partial n^0_b}{\partial \rho}$ as well as (f) $I_z=\int \rho d\rho dz
n^0_fz\frac{\partial n^0_b}{\partial z}$. The mean square radii in units of $a^2_{\rm ho}$ of the Bose and Fermi superfluids for the radial direction
are shown in (g) and ones for the axial direction in (h). The dashed lines represent the analytical results defined in Appendix,
and the discrete data indicate the numerical calculations for the coupled hydrodynamic equations at equilibrium.
The USTC experimental parameters \cite{yuping2018} are used.
}}
\end{figure}
\begin{figure}
\includegraphics[scale=0.27]{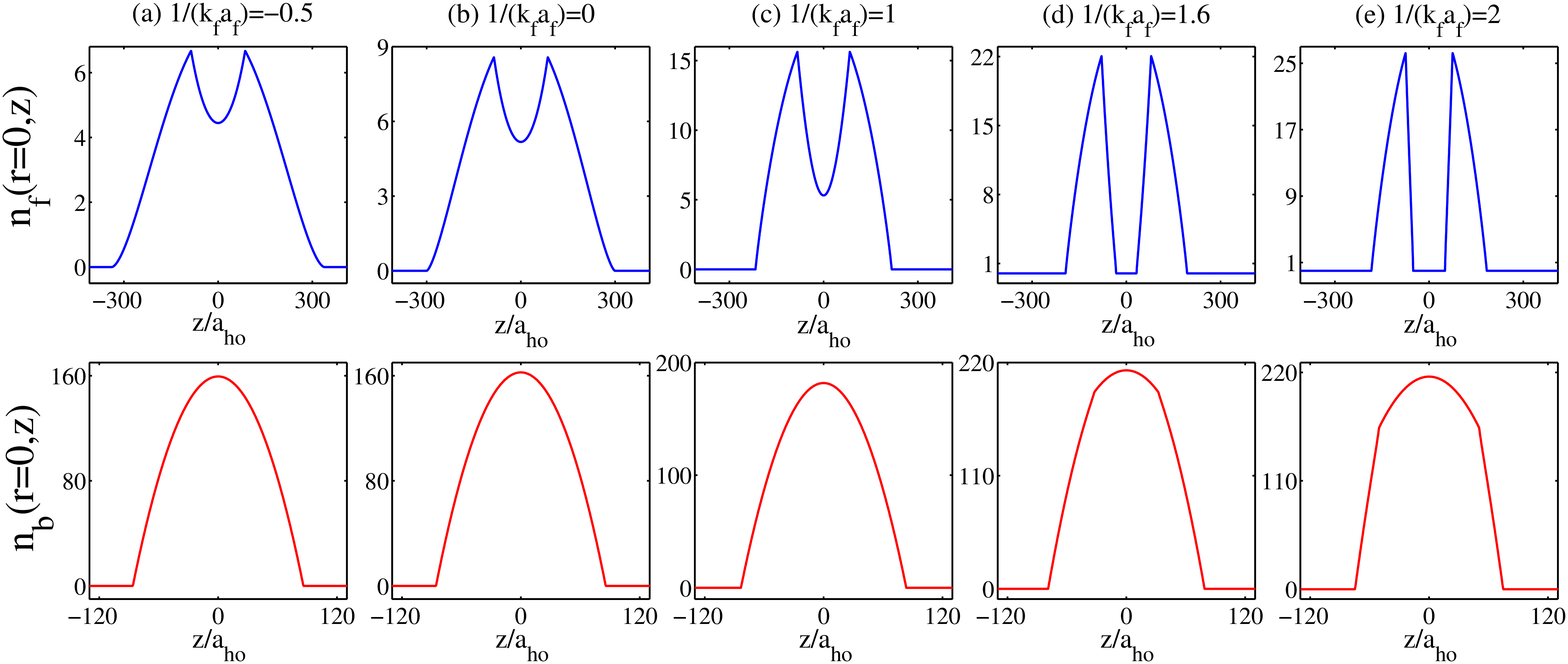}
\caption{\footnotesize{The axial density profiles at $r=0$ of the Fermi (upper panels)
and Bose (lower panels) superfluids in the BCS-BEC crossover interacting repulsively
in the USTC experimental setting. Both densities are given in units of $a^{-3}_{\rm ho}$.}}
\end{figure}

In order to test the approximations in Sec.\;IID,  we compare the analytical results with the numerical calculations in Fig.\;2.
We show the relevant integrals and the mean square radii as a function of the dimensionless parameter $1/(k_fa_f)$
for the USTC experimental setting (cigar-shaped case). The dashed lines are the analytical results defined in Appendix, and the discrete data are the results from solving the coupled
equilibrium hydrodynamic equations (\ref{bos}) and (\ref{fer}) numerically through a self-consistent iterative procedure \cite{wen2017}.
One can find that the analytical results agree well with the numerical
calculations, but showing obvious discrepancy when $1/(k_fa_f)>1$.
In order to explain such difference, in Fig.\;3 we plot the numerical results for the axial density profiles at $r=0$
for the fermions (upper panels) and bosons (lower panels) in the BCS-BEC crossover. One can find that
for a fixed Bose-Fermi interaction, as the interaction energy of
the Fermi superfluid decreases from the BCS side (Fig.\;3(a)) to the BEC regime (Fig.\;3(e)),
the depletion of the Fermi superfluid density in the center caused by the bosons becomes more and more pronounced, until it is completely in the BEC regime (see Fig\;3(d) and 3(e)).
Thus in the BEC side, the analytical approximation of substituting the Fermi density distribution in the integral region
by the noninteracting value in the center is unreliable.
It should be pointed that even through the Fermi density is zero in the center for the cases of Fig.\;3(d) and Fig.\;3(e),
the integrals for the spatial overlaps of these two densities still remain.
In addition, for the mean square radii of the Bose and Fermi superfluids shown in Fig.\;2(g) and 2(h),
compared with the analytical results (dashed lines) actually corresponding to the noninteracting Bose and Fermi superfluids, numerical results
(discrete data) show that the repulsive Bose-Fermi interaction has a very slight affect on the Fermi superfluids due to a larger number of
particles, while the radii of the Bose superfluids decreases obviously, suggesting that the bosonic cloud is compressed by the outer shell of
fermions.

In Fig.\;4, we first display the frequencies of the quadrupole oscillation modes of the Bose-Fermi
superfluid mixtures in the BCS-BEC crossover for different trap geometries.
The numerical results denoted by the open circles are obtained by solving the eigenvalue matrix (\ref{matrixr}) directly,
in which the relevant dimensionless parameters are determined by the numerical calculations for the Bose and Fermi superfluid density profiles
from the coupled equilibrium hydrodynamic equations \cite{wen2017}. By examining the signs of the eigenvectors, one
can identify the corresponding quadrupole modes of the Fermi and Bose superfluids.
In contrast, the analytical results are shown by the solid lines, which are calculated from the expressions (\ref{nmr}) combined with the analytical results for the dimensionless parameter and the mean square radii in Appendix. The positive (+) and negative (-) roots of the analytical expressions (\ref{nmr}), as the frequencies of the harmonic trap for fermions are larger than bosons,
actually correspond to the frequencies of the Fermi and Bose counterparts, respectively.
The frequencies of the quadrupole modes of the Bose and Fermi superfluid without interacting
are also plotted by dashed lines for comparison.

\begin{figure}
\includegraphics[scale=0.25]{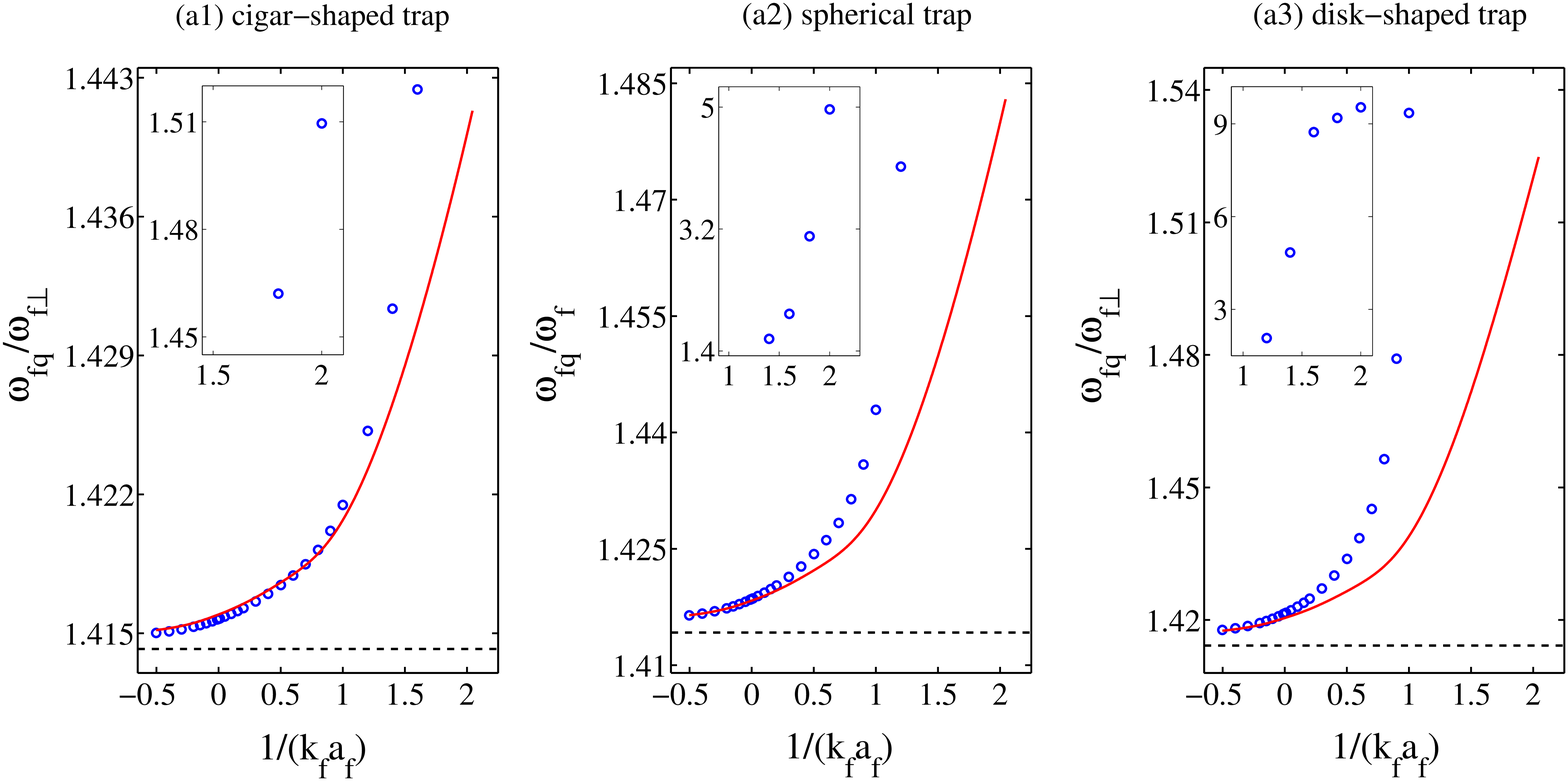}
\includegraphics[scale=0.25]{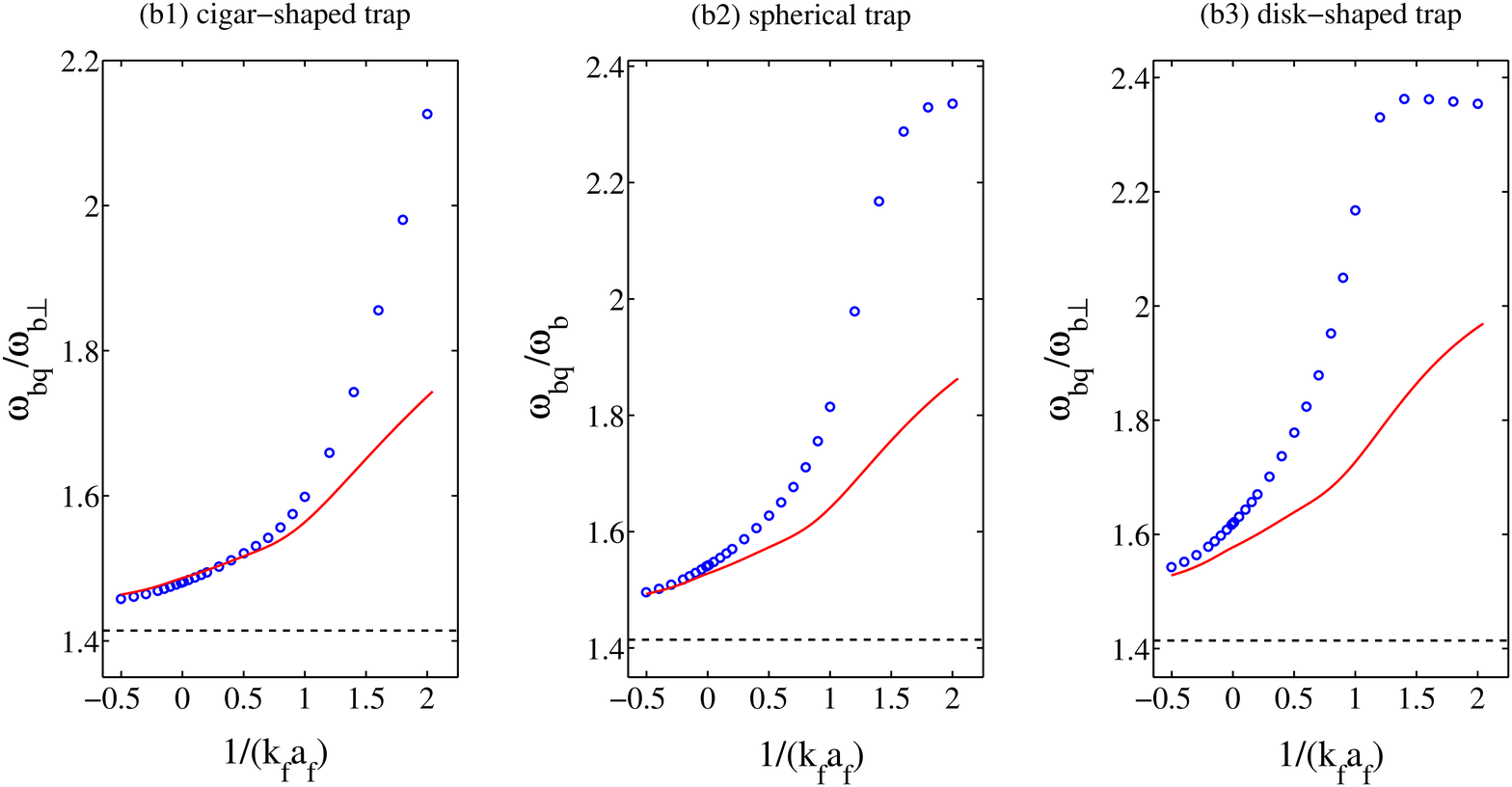}
\caption{\footnotesize{The frequencies of the quadrupole oscillation modes of the Fermi (upper panels) and
Bose (lower panels) superfluids with a repulsive Bose-Fermi interaction as a function of the dimensionless parameter $1/(k_fa_f)$
for different trap geometries: (1) cigar-shaped trap ($\lambda_b=0.04, \lambda_f=0.06$), (2) spherical trap ($\lambda_b=1, \lambda_f=1$),
and (3) disk-shaped trap ($\lambda_b=7, \lambda_f=11$). The solid lines represent the analytical results, while the numerical calculations are shown by the open circles, in which some larger values of the Fermi superfluids in the BEC regime are plotted in the corresponding insets. The dashed lines correspond to the Bose and Fermi superfluids alone.}}
\end{figure}

It is clearly seen that in the presence of a repulsive Bose-Fermi interaction, the frequencies of the quadrupole modes of the Fermi and Bose superfluids are all upshifted. Both the analytical and numerical results show that the upshifted values increase from the BCS side to the BEC regime, however, its behavior for the Fermi and Bose superfluids exhibits differently in the BEC regime.
The frequency of the quadrupole mode of the Fermi superfluid interacting with bosons increases monotonically from the BCS side to the BEC regime, and due to a larger number, the induced frequency shifts and their changes around the unitary limit are smaller than the Bose counterpart. However, the numerical results for the Fermi superfluid show a rapid increase for $1/(k_fa_f)>1$, some of which are plotted in the corresponding insets of Fig.\;4(a). Such significant rise can be understood by reexamining the equilibrium density profiles shown in Fig.\;(3). For a fixed repulsive Bose-Fermi interaction, as the interacting strength of the Fermi superfluid decreases from the BCS side to the BEC regime, the density of the Fermi superfluid overlapping with the bosons in the center of the trap decreases. In spite of the density reduction in the overlap region, the frequency shifts are actually sensitive to the spatial deviations of these two density distributions. From Fig.\;3(a) to Fig.\;3(e), the boundaries of the overlap regions become sharper and sharper, which result in more rapid increases of the overlap integrals and the effective Bose-Fermi coupling, and faster increases of the frequency shifts. However, the analytical results show an obvious slower increase, since the approximations of the analytical analysis underestimate these integrals (see Fig.\;2).

Differently from the fermionic counterpart, the numerical results for the frequency shifts of the Bose superfluid show non-monotonic increases from the BCS side to BEC regime for the spherical (Fig.\;4(b2)) and the disk-shaped (Fig.\;4(b3)) cases. Here to realize the traps from cigar-shaped to disk-shaped we increase the axial trapping frequencies, which actually also enhance the overlaps of the Bose and Fermi densities and the Bose-Fermi coupling effects. This is the reason why the frequency shifts in the disk-shaped case are largest and the discrepancy between the analytical and numerical results is most significant. We find that in the BEC regime $1/(k_fa_f)>1.5$ where the depletion of the Fermi superfluid is completely repelled by the bosons, although the overlap integrals for spatial derivations of the densities and the effective Bose-Fermi coupling increases monotonically, the frequency shifts of the Bose superfluids in the spherical (Fig.\;4(b2)) and disk-shaped (Fig.\;4(b3)) traps increasingly reach to a peak, then decrease slightly. Therefore one can find that the different increase speeds of the frequency shifts can be used to discriminate different equilibrium configurations of the Bose-Fermi superfluid mixtures in the BCS-BEC crossover.

\begin{figure}
\includegraphics[scale=0.25]{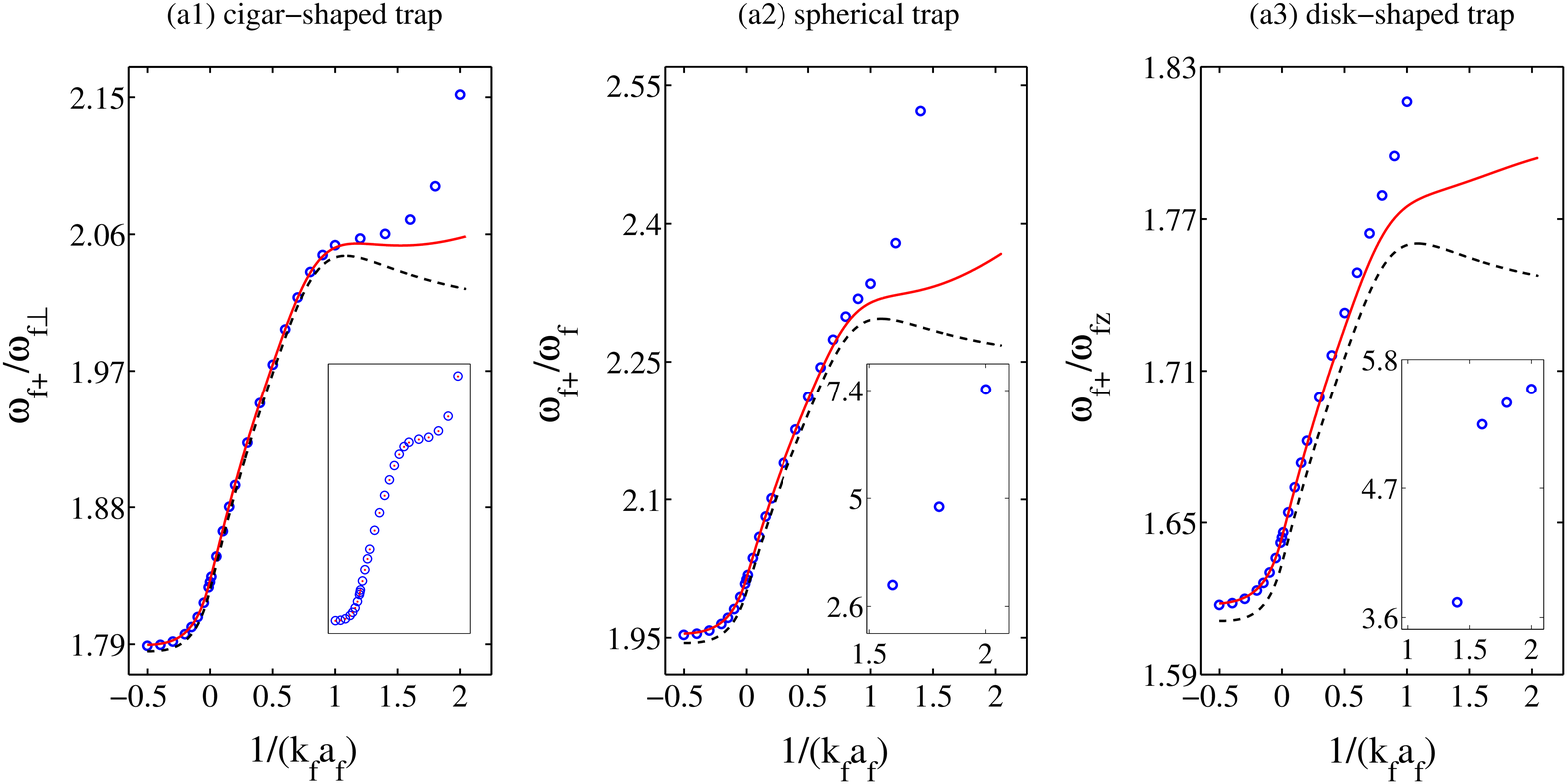}
\includegraphics[scale=0.25]{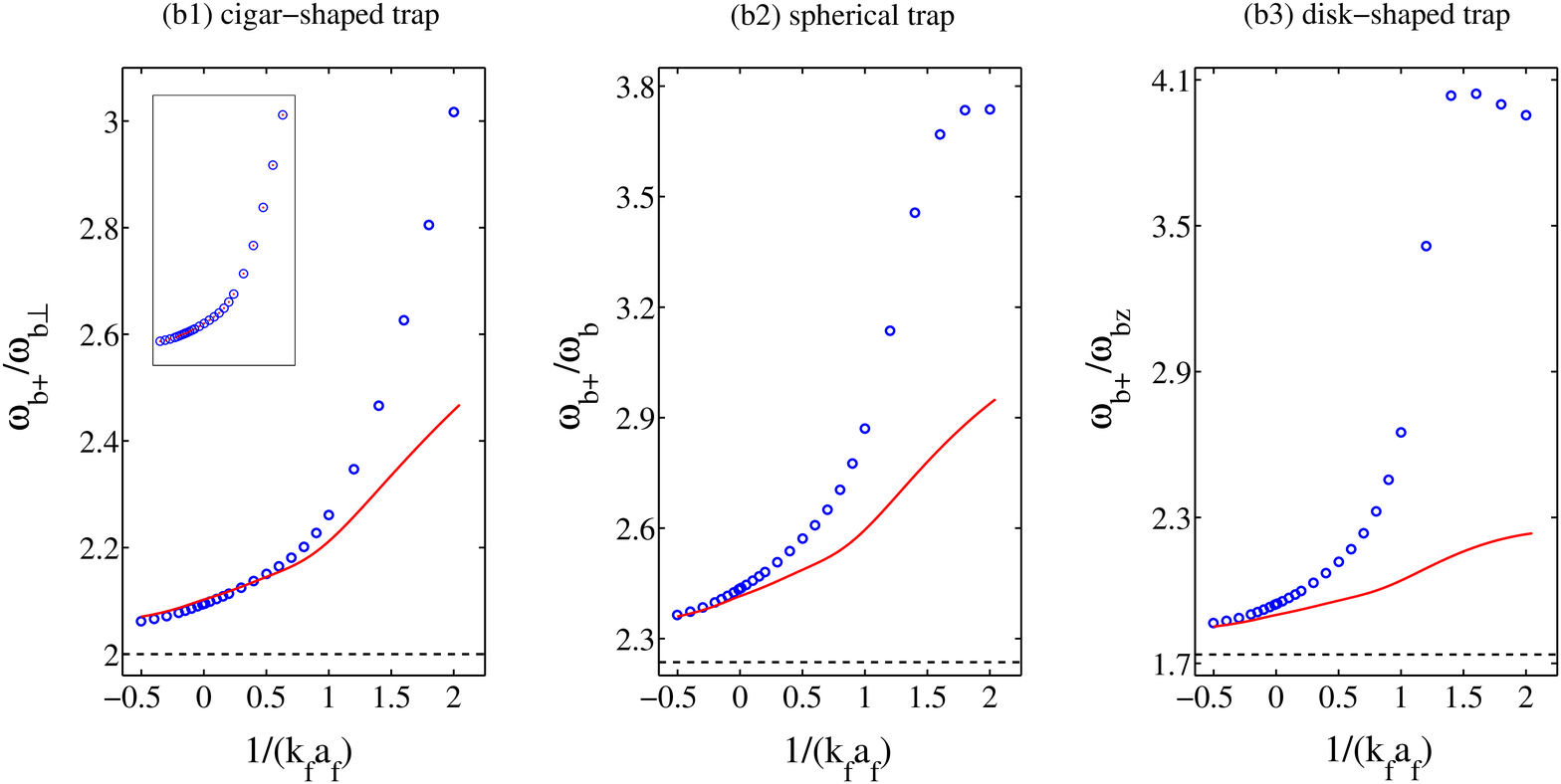}
\caption{\footnotesize{The frequencies of the radial oscillation modes of the (upper panels) Fermi and (lower panels)
Bose superfluids with a repulsive Bose-Fermi interaction as a function of the dimensionless parameter $1/(k_fa_f)$
for different trap geometries. The solid lines represent the analytical results, and the open circles correspond to the
numerical calculations, in which some larger values of the Fermi superfluids in the BEC regime are plotted in the insets
of (a2) and (a3). The insets of (a1) and (b1) correspond to the numerical results for the radial modes
(open circles) and the radial (transverse) breathing modes (dots). The dashed lines show the Bose and Fermi superfluids alone.
}}
\end{figure}

\begin{figure}
\includegraphics[scale=0.25]{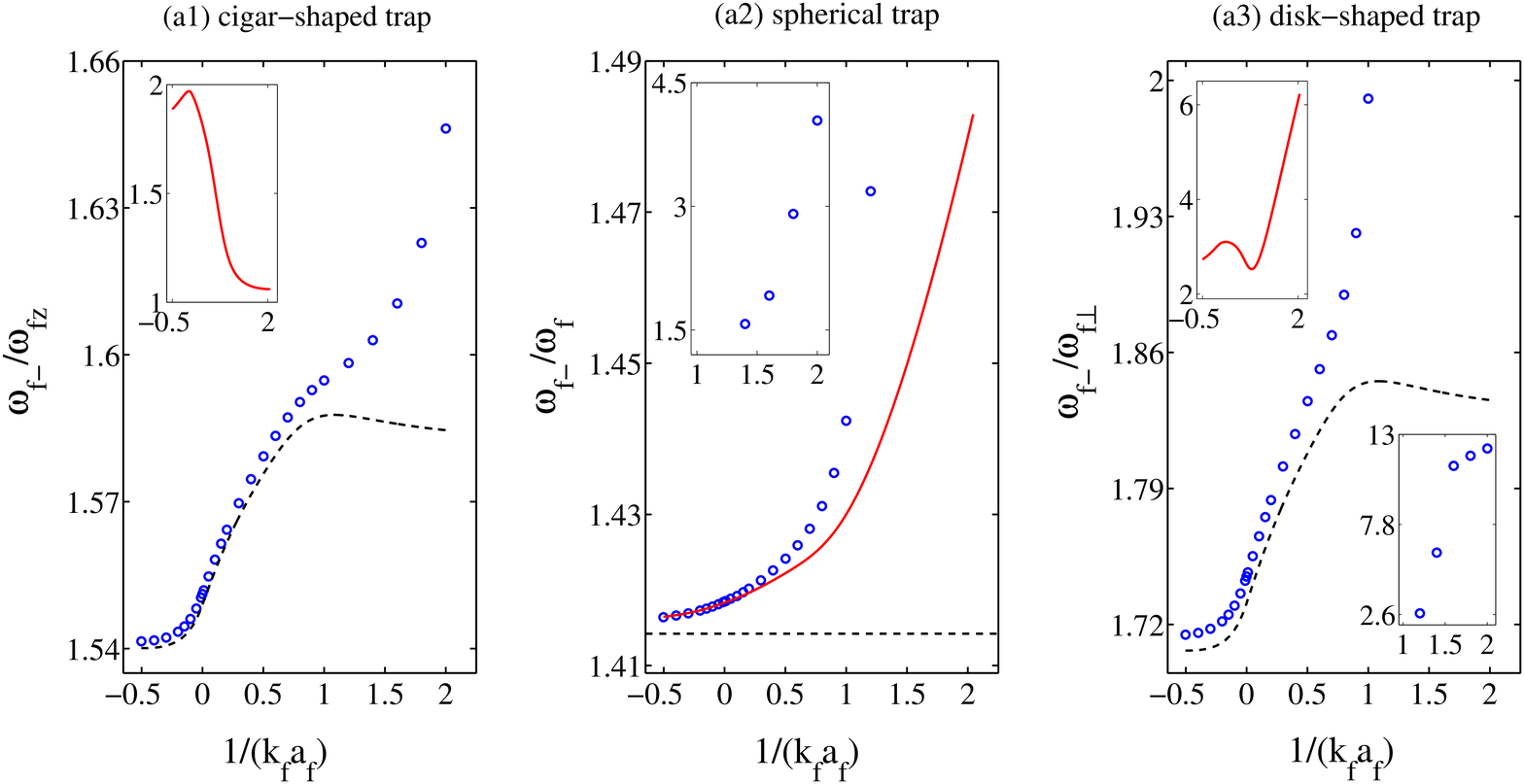}
\includegraphics[scale=0.25]{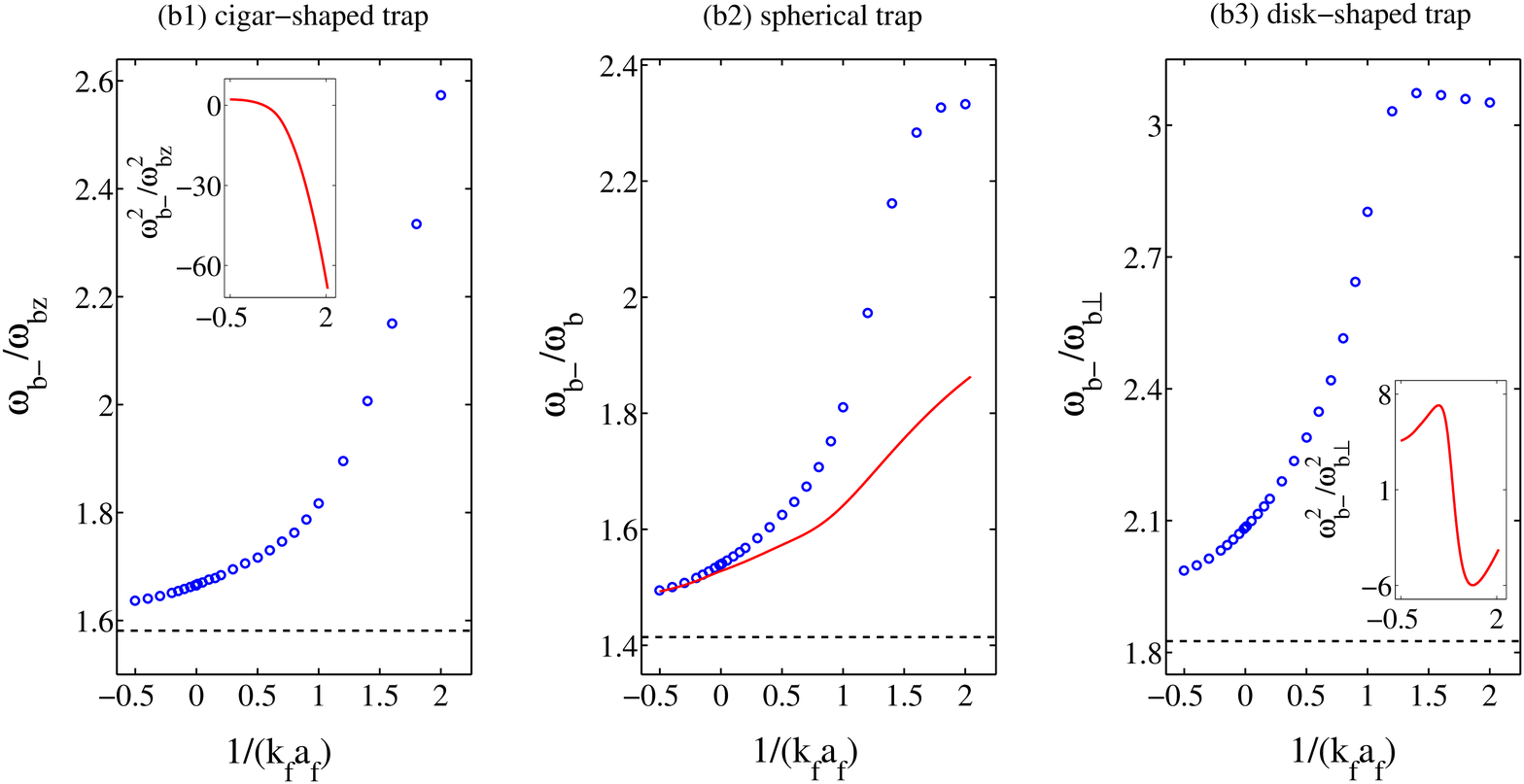}
\caption{\footnotesize{The frequencies of the axial oscillation modes of the (upper panels) Fermi and (lower panels) Bose superfluids
with a repulsive Bose-Fermi interaction as a function of the dimensionless parameter $1/(k_fa_f)$ for the different trap geometries.
The numerical calculations are shown by the open circles, and some larger values of the Fermi superfluids in the BEC regime
are plotted in the corresponding insets of (a2) and (a3). The analytical results are denoted by the solid lines, and
the cases in the anisotropic traps (i.e. (a1) and (a3) for the Fermi superfluid and (b1) and (b3) for the Bose superfluid)
are shown in the corresponding insets. The dashed lines correspond to the Bose and Fermi superfluids alone.
}}
\end{figure}
The analytical (solid lines) and numerical (open circles) results for the frequencies of the radial modes of the Bose and Fermi
superfluids as a function of the dimensionless parameter  $1/(k_fa_f)$ are shown in Fig.\;5. The results for noninteracting Bose and Fermi superfluids (dashed lines) are also plotted for comparison. In contrast to the quadrupole mode in the absence of the Bose-Fermi interaction,
the radial mode frequency of the Fermi superfluid shows a non-monotonical variation as a function of the dimensionless parameter in Fig.\;5(a), which are also relevant to the trap anisotropy. As discussed in Sec.\;IIC, the radial modes of a single superfluid in a highly elongated trap reduce to the radial (transverse) breathing mode  \cite{vic1996,hei2004,huang2018}, which is featured by the radii in the transverse plane oscillating in phase with each other without axial excitation. In the insets of Fig.\;5(a1) and 5(b1), we compare the numerical results for the frequencies of the radial modes (open circles) of the Fermi and Bose superfluids in the cigar-shaped traps to those of the radial breathing modes (dots), respectively. The radial breathing modes are calculated by numerically solving the eigenvalue matrix (\ref{matrixr}) and distinguished
by the signs of the eigenvectors. One can find the frequencies of these two modes are quite the same, which implies that in the presence of the Bose-Fermi interaction such symmetry of the system is still preserved.

In Figs.\;6(a) and 6(b), we show the frequencies of the axial modes of
the Fermi and Bose superfluids, respectively, as a function of the dimensionless parameter  $1/(k_fa_f)$  for different trap geometries.
For the cases in the cigar-shaped and disk-shaped traps, we plot the analytical results for the frequencies of the Fermi superfluids separately in the insets of Fig.\;6(a1) and Fig.\;6(a3), and the squares of the frequencies for the Bose counterpart in the insets of Fig.\;6(b1) and Fig.\;6(b3), respectively. The reason for the failure of the analytical expressions for the axial modes in the highly-anisotropic traps is that from the explicit expressions (\ref{nmm}) and the eigenvalue matrix (\ref{matrixm}), one can find the elements $A^{\pm}_{12,21}$ characterizing the coupling of the Bose and Fermi superfluids, are not only determined by the dimensionless parameters as the quadrupole mode, but also rely on the ratio $M^{\pm}_{b,f}$ of the radial excitation amplitude to axial excitation amplitude. We find that for the cases of the axial modes in the highly-anisotropic traps, $M^{-}_{b,f}$ leads to the coupling terms $A^{-}_{12,21}$ comparable to the non-coupling terms $A^{-}_{11,22}$, however, the analytical expressions are only justified for a weak coupling, i.e. the coupling part is much smaller than the non-coupling one.
In anisotropic traps, the radial and axial modes result from the coupled quadrupole and monopole modes \cite{hei2004}. In spherical traps without the coupling \cite{sstri1996}, the radial mode is also defined by the monopole mode and the axial mode recovers to the quadrupole mode. By comparing the frequencies of the axial modes of the Fermi (Fig.\;6(a2)) and Bose (Fig.\;6(b2)) superfluids in the spherical trap with
those of the quadrupole modes of the Fermi (see Fig.\;4(a2)) and Bose (see Fig.\;4(b2)), respectively, we find that they are the quite same, expect for the numerical value for the case of the Fermi superfluid at $1/(k_fa_f)=2$ due to large mixing of the eigenvetors of the axial modes.

\section{Conclusion}

Since the realization of a mixture of Bose and Fermi superfluids in the BCS-BEC crossover in ultracold atoms
the investigation of the properties of collective modes will be of particular interest, which serve as a powerful tool
to understand the physics of many-body systems. In this work, we study the quadrupole, radial and axial oscillation modes of the Bose-Fermi superfluid mixtures in the BCS-BEC crossover in anisotropic traps by using the coupled hydrodynamic superfluid equations
and the scaling theory for the coupled system. The analytical analysis for the frequencies of collective oscillation modes are presented,
and the valid regimes of the approximations are demonstrated by the numerical calculations in currently experimentally feasible setups.
We find that for a repulsive Bose-Fermi interaction, the frequencies of the quadrupole, radial and axial modes of the Bose
and Fermi superfluids are all upshifted, and the frequency shifts of the Fermi superfluid are smaller than the bosonic counterpart
due to a larger number of particles. However, for a fixed repulsive Bose-Fermi interaction, as we pass from the BCS side to the BEC regime
where the interaction energy of the Fermi superfluid decreases, the frequency shifts of the collective oscillation modes increase,
especially for the Fermi superfluid in the BEC regime, and the change speeds of the frequency shifts in different superfluid regimes can be
used to characterize different density profiles of the Bose-Fermi superfluid mixtures at groundstates. We also find that compared with the quadrupole and axial modes, the frequency shifts of the radial modes of the Bose superfluids are the most significant for the same case, which should be easily measured by experiments \cite{huang2018}.

The theoretical results obtained here from the scaling method for a coupled system may provide a useful reference for future experiments
on the collective excitations of Bose-Fermi superfluid mixture in the BCS-BEC crossover, and on the other hand, the availability of Bose-Fermi superfluid mixtures with a long-lived lifetime can provide a unique probability for testing the scaling theory of a coupled system, which
has been confirmed experimentally for a single superfluid. However, the issue of collective oscillation modes of Bose-Fermi superfluid mixtures
in the BCS-BEC crossover is far from simplicity as we study. Under the TF approximation, we ignore the density gradient corrections,
which prevent the densities from changing abruptly. In the vicinity of sharp edges, the density will be smoothed out by including the
corrections. So the overlap integrals of the spatial derivations of the Bose and Fermi density profiles have been overestimated,
as well as the rapid increase of the frequency shifts, but the TF approximation is still expected to provide qualitatively correct results.
Furthermore, we ignore all decay processes, which may be resulted from normal components, single particle excitations, and the
nonlinear coupling of high amplitude oscillations, and should be further considered theoretically and experimentally.

\acknowledgments
This work is supported by the NSFC under Grant No. 11105039 and the Fundamental Research Funds
for the Central Universities (No. 2019B21214 and No. 2017B18014).

\appendix*
\section{Explicit expressions for the integrals and dimensionless parameters}

The explicit expressions for the integrals $I_{\alpha\beta}$ and $I_{\alpha}$ with $\alpha,\beta=\rho(\perp),z$ are given by, respectively,
\begin{eqnarray}\label{allinterr}
& & I_{\alpha\beta}=\int \rho d\rho dz\frac{\partial n^0_f}{\partial r_\alpha}\alpha\beta \frac{\partial n^0_b}{\partial r_\beta}
=\frac{i_{\alpha\beta}}{7\pi}\mu_bN_bC_{\alpha\beta},\;\;\; i_{\rho\rho}=4i_{\rho z(z\rho)}=\frac{8}{3}i_{zz}=8, \\
& & C_{\alpha\beta}=(\frac{\partial n^{00}_f}{\partial \mu_f})_{{\bf r}=0}\left[
g_{bf}(\frac{\partial n^{00}_f}{\partial \mu_f})_{{\bf r}=0}\frac{m_f\omega^2_{f\beta}}{m_b\omega^2_{b\beta}}-1\right]\left(\frac{g_{bf}}{g_b}-\frac{m_f\omega^2_{f\alpha}}{m_b\omega^2_{b\alpha}}\right)\\
& & I_{\alpha}=\int \rho d\rho dz \frac{\partial n^0_b}{\partial r_\alpha}\alpha n^0_f=n^{00}_f(0)\frac{i_{\alpha}N_b}{\pi}C_{\alpha},\;\;\;\;\;\;\; i_{\rho}=2i_z=1,\\
& & C_{\alpha}=g_{bf}(\frac{\partial n^{00}_f}{\partial \mu_f})_{{\bf r}=0}\frac{m_f\omega^2_{f\alpha}}{m_b\omega^2_{b\alpha}}-1,
\end{eqnarray}
and the dimensionless parameters are given by
\begin{eqnarray}\label{alldimension}
& & B_{\alpha\beta}=\frac{2\pi g_{bf}}{N_bm_b\omega^2_{b\alpha}{\langle R_\alpha^2\rangle}_b}I_{\alpha\beta}=g_{bf}i_{\alpha\beta}C_{\alpha\beta},\;\;\;\; i_{\rho\rho}=4i_{\rho z}=2i_{z\rho}=\frac{4}{3}i_{zz}=4,\\
& & F_{\alpha\beta}=\frac{2\pi g_{bf}}{N_fm_f\omega^2_{f\alpha}\langle R_\alpha^2\rangle_f}I_{\beta\alpha}=g_{bf}\frac{5\gamma+2}{7\gamma}\frac{\mu_bN_b}{\mu_fN_f}i_{\alpha\beta}C_{\beta\alpha},\\
& & F_{\alpha}=\frac{2\pi g_{bf}}{N_fm_f\omega^2_{f\alpha}{\langle R_\alpha^2\rangle}_f}I_{\alpha}=g_{bf}\frac{5\gamma+2}{2}\frac{N_b}{N_f}(\frac{\partial n^{00}_f}{\partial \mu_f})_{{\bf r}=0}C_{\alpha},
\end{eqnarray}
where $\langle R_\alpha^2\rangle _b=(1/N_b)\int d{\bf r}\;r_\alpha^2 n^{00}_b({\bf r})
=i_{\alpha}R^2_{b\alpha}/7$ are the mean square radii of the Bose superfluid distributions in the ($\alpha=\perp,z$) direction,
and $\langle R_{\alpha}^2\rangle _f=(1/N_f)\int d{\bf r}\;r_\alpha^2n^{00}_f({\bf r})=i_{\alpha}\gamma R^2_{f\alpha}/(2+5\gamma)$
for the Fermi superfluid with $i_{\rho}=2i_z=2$. The equilibrium TF radii for the Bose superfluid alone
are given by $R_{b\alpha}=\sqrt{2\mu_b/(m_b\omega^2_{b\alpha})}$ and $R_{f\alpha}=\sqrt{2\mu_f/(m_f\omega^2_{f\alpha})}$ for the Fermi superfluid alone.


\end{document}